\documentclass[a4paper,fleqn,usenatbib]{mnras}

\usepackage{graphicx}	% Including figure files
\usepackage{amsmath}	% Advanced maths commands
\usepackage{amssymb}	% Extra maths symbols
\usepackage{multicol}        % Multi-column entries in tables
\usepackage{bm}		% Bold maths symbols, including upright Greek
\usepackage{pdflscape}	% Landscape pages

\usepackage[T1]{fontenc}
\usepackage{ae,aecompl}

\usepackage{txfonts}
\title[Gaseous dwarf-dwarf interactions]{Local Volume TiNy Titans:\\ Gaseous Dwarf-Dwarf Interactions in the Local Universe}

\author[S. Pearson et al. ]{
Sarah Pearson$^1$\thanks{Contact e-mail: \href{mailto:spearson@astro.columbua.edu}{spearson@astro.columbia.edu}},
Gurtina Besla$^2$, 
Mary E. Putman$^1$,
Katharina A. Lutz$^{3,4}$, 
\newauthor
Ximena Fern\'{a}ndez$^{5,6}$, 
Sabrina Stierwalt$^{7}$, 
David R. Patton$^8$, 
\newauthor
Jinhyub Kim$^9$,  
Nitya Kallivayalil$^{10}$, 
Kelsey Johnson$^{10}$, 
Eon-Chang Sung$^{11}$
\\
$^1$ Department of Astronomy, Columbia University, New York, NY 10027, USA\\
$^2$ Department of Astronomy, University of Arizona, Arizona, USA\\
$^3$ Centre for Astrophysics and Supercomputing, Swinburne University of Technology, Hawthorn, VIC, Australia\\
$^4$ ATNF, CSIRO Astronomy and Space Science, PO Box 76, Epping, NSW 1710, Australia\\
$^5$ Department of Physics and Astronomy, Rutgers, The State University of New Jersey, NJ 08854-8019 USA\\
$^6$ NSF Astronomy and Astrophysics Postdoctoral Fellow\\
$^{7}$ National Radio Astronomy Observatory, 520 Edgemont Rd, Charlottesville, VA 22903, USA\\
$^8$ Department of Physics \& Astronomy, Trent University, 1600 West Bank Drive, Peterborough, ON K9J 7B8, Canada\\
$^9$ Department of Astronomy and Yonsei University Observatory, Yonsei University, Seoul 120-749, Korea\\
$^{10}$ Department of Astronomy, University of Virginia, Virginia, USA\\
$^{11}$ Korea Astronomy and Space Science Institute, 776 Daedeok-daero, Yuseong, Daejeon 305-348, Korea\\
}
\date{\today}

\pubyear{2016}

\begin{document}
\label{firstpage}
\pagerange{\pageref{firstpage}--\pageref{lastpage}}
\maketitle

% Abstract of the paper
\begin{abstract}
In this paper, we introduce the Local Volume TiNy Titans sample (LV-TNT), which is a part of a larger body of work on interacting dwarf galaxies: TNT (\citealt{stierwalt15}). This LV-TNT sample consists of 10 dwarf galaxy pairs in the Local Universe ($<$ 30 Mpc from Milky Way), which span mass ratios of M$_{*,1}$/M$_{*,2}$ $<$ 20, projected separations $<$ 100 kpc, and pair member masses of log(M$_*$/M$_{\odot}$) $<$ 9.9. All 10 LV-TNT pairs have resolved synthesis maps of their neutral hydrogen, are located in a range of environments and captured at various interaction stages. This enables us to do a comparative study of the diffuse gas in dwarf-dwarf interactions and disentangle the gas lost due to interactions with halos of massive galaxies, from the gas lost due to mutual interaction between the dwarfs. We find that the neutral gas is extended in the interacting pairs when compared to non-paired analogs, indicating that gas is tidally pre-processed. Additionally, we find that the environment can shape the HI distributions in the form of trailing tails and that the gas is not unbound and lost to the surroundings unless the dwarf pair is residing near a massive galaxy. We conclude that a nearby, massive host galaxy is what ultimately prevents the gas from being reaccreted. Dwarf-dwarf interactions thus represent an important part of the baryon cycle of low mass galaxies, enabling the ``parking'' of gas at large distances to serve as a continual gas supply channel until accretion by a more massive host.

\end{abstract}


%%%%%%%%%%%%%%%%% BODY OF PAPER %%%%%%%%%%%%%%%%%%

\section{Introduction} \label{sec:intro}
In the current $\Lambda$CDM paradigm interactions and mergers between galaxies in the Universe are thought to play an important role in their evolution. Interactions between massive galaxies have been studied extensively both observationally and theoretically. Through these studies, we know that the interactions induce morphological changes (e.g.  \citealt{toomre72}, \citealt{propris07}, \citealt{ellison10}, \citealt{casteels14}), can trigger starbursts (e.g. \citealt{hernquist89}, \citealt{scudder12}, \citealt{patton13}, \citealt{davies15}), gas inflows and also AGN activity (e.g. \citealt{sanders88}, \citealt{ellison11}, \citealt{silverman11}, \citealt{satyapal14}). Recently, \citet{seng15} studied the tidal forces acting on the gas in prograde, low-velocity flyby interactions, which showed that large fractions of the gas can be displaced beyond the optical disks of massive spirals. However, very little is known observationally and theoretically about interactions between the most prevalent type of galaxies at all redshifts, dwarf galaxies (\citealt{bing98}; \citealt{kara13}). Dwarf galaxies differ from their more massive counterparts, as their dark matter to baryonic matter fraction tends to be larger, they are much fainter (10-10$^5$ times fainter than Milky Way (MW) type galaxies) and they are inefficient at forming stars from their large gas reservoirs (e.g. \citealt{blanton01}, \citealt{robertson08}). It is therefore unclear whether the same processes observed in massive galaxies (e.g. their merger sequence and their displacement of gas with interaction) scale down to the dwarf regime. Since dwarf galaxies dominate the galaxy population of the Universe at all times, the interactions and mergers between them occur more frequently in a given volume than for massive galaxies (\citealt{fak10}, \citealt{deason14}) and they play a crucial role in the hierarchical build up of dark matter and stellar halos. 
Thus, studying their mutual interaction does not only shed light on the merger sequence of dwarfs, but their interactions also play a key role in how more massive galaxies are fed by accretion, as dwarf galaxies are often captured as pairs or groups (e.g. \citealt{wetzel15}). Pre-processing of gas by tidal interactions prior to infall can drastically increase the efficiency of gas lost to the halos of the more massive galaxies (see e.g. \citealt{besla10}, \citealt{salem15}), hence understanding the baryon cycle in dwarf galaxy interactions provides insight into how the baryon cycle of more massive galaxies is acquired. 

Recently, \citet{stierwalt15} carried out a systematic study of 104 dwarf galaxy pairs in a wide range of interaction stages and environments, selected from the Sloan Digital Sky Survey (the TiNY Titans Survey). The pairs in their sample all have projected separations $<$ 50 kpc, velocity separations $<$ 300 km s$^{-1}$ and pair member masses between 7 $<$ log(M$_{*}$/M$_{\odot}$) $<$ 9.7. Interestingly, they find that the star formation rates of the dwarf galaxy pairs averaged over all pairs are enhanced with decreasing pair separation, which is also found for more massive galaxies (e.g. \citealt{patton13}). However, when they investigated the gas fractions of their dwarf pairs using single-dish Arecibo Telescope and Green Bank Telescope data, only pairs within 200 kpc of a massive host galaxy showed signs of gas depletion, whereas the dwarf pairs farther than 200 kpc from a massive host had large atomic gas fractions ($f_{\text{gas}}$ $>$ 0.6, where $f_{\text{gas}}$ = $1.4 M_{\text{HI}}/(1.4 M_{\text{HI}} + M_*)$). 

Similarly, \citet{bradford15} investigated the baryon content of low mass galaxies using single-dish Arecibo, Green Bank and remeasured ALFALFA data. They found that non-isolated low mass galaxies (in the vicinity of a massive host) had a higher scatter towards lower atomic gas fractions, when compared to isolated galaxies of similar masses\footnote{Although their median atomic gas fractions were very similar for non-isolated and isolated low mass galaxies ($f_{\text{gas}}$ $=$ 0.81 $\pm$ 0.13 and $f_{\text{gas}}$ $=$ 0.82 $\pm$ 0.16, respectively).} (see their Figure 4). Additionally, only non-isolated galaxies were gas-depleted as defined by a -1 dex deviation from the gas mass vs stellar mass fit to all data (\citealt{bradford15}, Figure 5). Lastly, they found that no isolated low mass galaxies had gas fractions lower than $f_{\text{gas}}$ $<$ 0.3, which sets an upper limit on the amount of gas that can be consumed by star formation or removed by outflows or tides.

The studies by \citet{stierwalt15} and \citet{bradford15} (see also \citealt{geha06}, \citealt{geha12}) all indicate that environment plays a key role in quenching dwarf galaxies and dwarf galaxy pairs, and that dwarfs and dwarf pairs in an isolated environment have high atomic gas fractions. Recently, \citet{davies15} showed that the fraction of passive, non star forming galaxies is higher at all stellar masses for interacting pairs and groups of galaxies, than for galaxies in the field, which introduces an interesting question about the role of galaxy interactions in quenching at the low mass end.

However, a key issue is that the gas fractions estimated from single-dish data could also include gas that sits in the outskirts of the galaxies, which can have been removed by tidal forces, ram pressure stripping or blown out from star formation. To understand whether this gas is unbound and enriches the intergalactic medium (or circumgalactic medium of a host galaxy), if it falls back and refuels the star formation in the dwarfs or if it resides within the dwarfs themselves, we must resolve the diffuse component of the atomic gas. As the tidal structures can reach HI column densities as low as N(HI) $\sim 10^{18}$ cm$^{-2}$ (e.g. for the LMC: \citealt{putman03}), mapping these structures at large distances, beyond the Local Volume, is difficult with existing instruments. 

The Magellanic System (MS) is currently the best template for an ongoing nearby dwarf-dwarf interaction. Evidence for this interaction is most convincingly seen in the extended HI distribution; this includes a bridge of gas connecting the two galaxies and two streams of gas that span $>$ 100$^{\text{o}}$ of sky (\citealt{putman03}). \citet{besla12} showed the importance of tidally pre-processing the gas prior to infall in shaping the tail of the MS. This scenario was supported by \citet{salem15} as they showed that ram pressure stripping is not sufficient to explain the amount of gas in the MS. Furthermore, \citet{salem15} showed that the truncation in the LMC's HI disk due to its motion through the Milky Way halo, can be used to constrain the Milky Way's halo density. However, having more examples than the MS is crucial for establishing the stage at which ram pressure stripping becomes important for removing gas and the role of minor mergers of groups in feeding the circum galactic medium (CGM) of galaxies like the Milky Way. 

Examples of Local Volume dwarf samples include \citet{odewahn94} and \citet{wilcots04}. While \citet{odewahn94} found that nearly all Magellanic Irregulars have nearby companions, \citet{wilcots04} showed that many of these companions might be chance projections. Additionally, the \citet{wilcots04} study of 13 Magellanic Irregulars found that the asymmetric morphology in the HI profiles of Magellanic Irregulars with gaseous companions did not seem to differ much from Magellanic Irregulars without gaseous companions. In this paper, we are selecting examples from the literature that are likely to be interacting, and therefore expect to find significant differences between our pairs and isolated analogs. In particular, we investigate the tidal and environmental effects on the diffuse gas in 10 Local Volume dwarf galaxy pairs with low relative velocities within 30 Mpc of the Milky Way. 

This paper is a part of a larger body of work (the TiNy Titans Survey (TNT)) and represents the Local Volume TNT sample (LV-TNT). Here we quantify the HI morphologies, HI surface density profiles and the fraction of neutral gas that resides outside vs inside the dwarf galaxies to understand the gas removal process and baryon cycle in these systems. The 10 pairs represent a diverse sample of dwarf galaxy interactions, as they are located at different distances from massive galaxies and are captured in various interaction stages. The diversity of the sample enables us to ask: what is more important - environment or dwarf-dwarf interactions in removing gas to large radii? How much material can be removed in this way and does this material remain bound to the pairs?  

The paper is structured as follows: In Section \ref{sec:sample} we describe the selection process leading to our sample, and present the properties of 10 dwarf galaxy pairs. In Section \ref{sec:results}, we show the results of our analyses of the diffuse gas from the 10 dwarf pairs. We discuss the results of our study in Section \ref{sec:discussion}, and we conclude in Section \ref{sec:conclusion}.

%----------------------------------------------------Sample-----------------------------------------------------------------
\section{Dwarf pair selection and sample}\label{sec:sample}
In this Section, we present the sample of 10 Local Volume dwarf galaxy pairs (see Figure \ref{fig:sample}). We describe the sample selection in Section \ref{sec:selec}, we describe how properties of the dwarf galaxies were estimated in Section \ref{sec:prop} and we discuss each dwarf pair in detail in Appendix~\ref{sec:samp}.

\subsection{Sample Selection}\label{sec:selec}
Our goal is to investigate the gas distributions around dwarf galaxy pairs. Therefore, we are only interested in Local Volume, interacting dwarf galaxies where the HI structure can be resolved to sufficiently explore the gas beyond the optical extents of the dwarfs. To enforce these criteria, we require that the projected separation, $R_\text{sep}$, between each pair is $<$ 100 kpc (in practice, only one of our targets has a $R_\text{sep}$ $>$ 50 kpc (see Figure \ref{fig:mass})), the velocity separation, $v_\text{sep}$, is $<$ 300 km s$^{-1}$ and that all pairs are within 30 Mpc of the Milky Way. The limits on $R_\text{sep}$ and $v_\text{sep}$ are similar to those in the TiNY Titans sample (\citealt{stierwalt15}) and to those known to identify interactions between more massive galaxies (e.g. \citealt{patton13}). 
We set an upper stellar mass limit for each dwarf galaxy in our sample at $M_{*}$ $< 10^{10}$ M$_{\odot}$, and we require that the dwarf galaxy pairs must have HI maps with an outer column density limit of N(HI) $<$  7 $\times$ 10$^{19}$ atoms cm$^{-2}$.

After applying these cuts to pairs found through an extensive literature search, we have 10 Local Volume dwarf galaxy pairs, which we present in Section \ref{sec:sample} and list in Tables \ref{tab:local}, \ref{table:prop} and \ref{table:HI}. We preferentially selected pairs where evidence of interaction has been presented in the literature, and are therefore biased to systems that are connected by bridges and show signs of interaction through tidal features. This sample can be expanded in the future with further HI maps of existing and newly identified pairs. As the number of dwarf pairs in the Local Volume is observationally and theoretically uncertain, assessing the completeness of our sample is difficult. However, we stress that the goal of this study is not to present a comprehensive census of dwarf pairs in the Local Volume, rather we chose 10 targets that range in projected separation, velocity, mass ratio and environment to probe potential signatures of dwarf-dwarf interactions or environmental effects.

The resolved HI maps for each of our targets are shown in Figure \ref{fig:sample}, and the details of the observations are listed in Table \ref{table:HI}. The mass ratios, relative line of sight velocities and pair separations of all pairs are shown in Figure \ref{fig:mass}. These are all within the velocity separation used in the \citet{stierwalt15} sample, however we include one pair that is at a projected separation of $\sim$ 100 kpc, and four pairs that have stellar mass ratios 10 $<$ $(M_1/M_2)^*$ $<$ 20 (the \citet{stierwalt15} limit is $R_\text{sep}$ $<$ 50 kpc and $(M_1/M_2)^* <$ 10).

\subsection{Properties}\label{sec:prop}
\begin{figure*}
\centerline{\includegraphics[width=\textwidth]{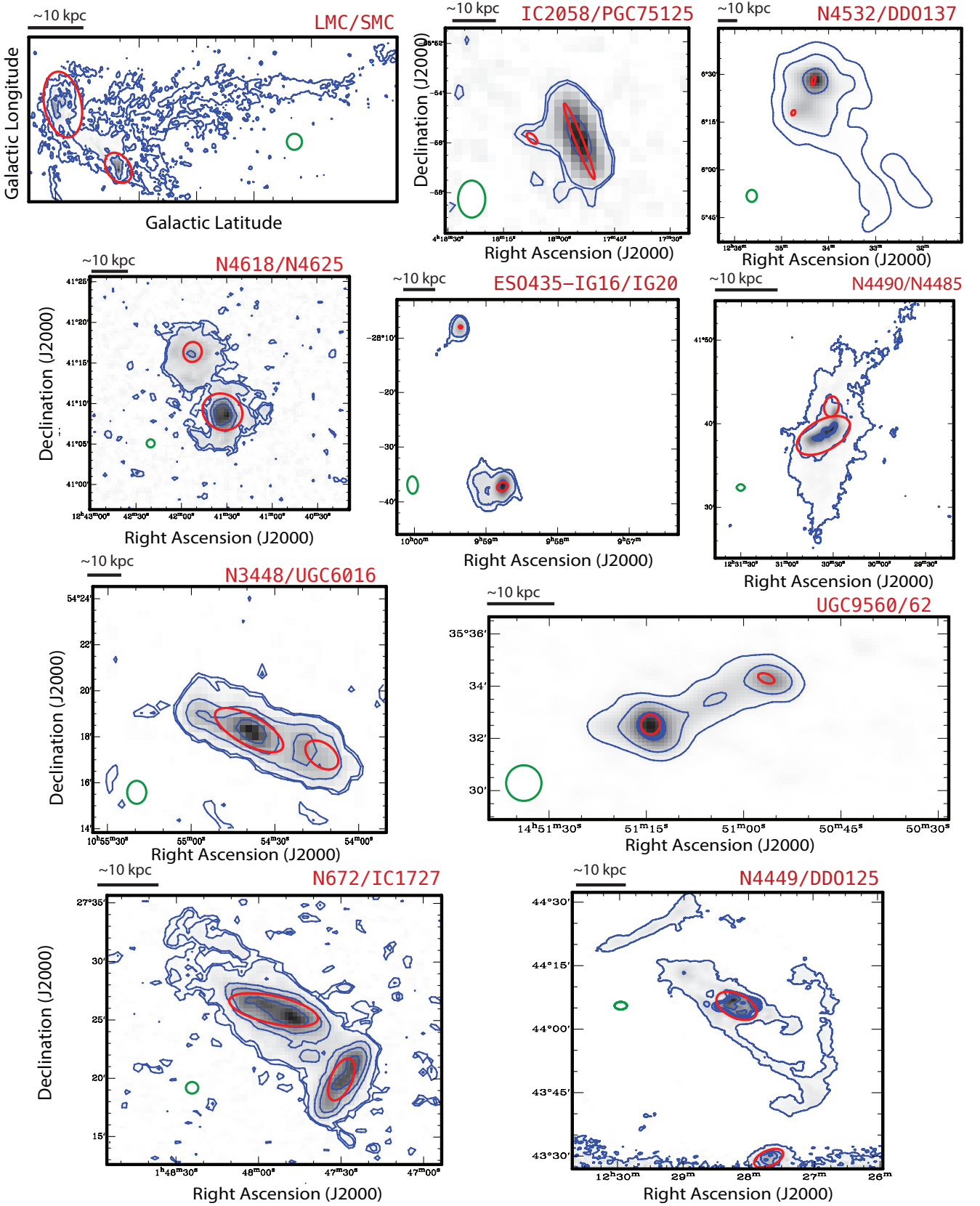}}
\caption{HI contours (blue) of all LV-TNT pairs in order of decreasing tidal index, with the 2MASS ellipses in (red) in addition to the beam sizes (green). All N(HI) are listed in units of $\times$ 10$^{20}$ atoms cm$^{-2}$. 
\textbf{1st row:} N(HI)$_{\text {LMC}}$ = 0.1, 1.0, 10.0. N(HI)$_{\text {IC 2058}}$ = 0.7, 1.0, 10. N(HI)$_{\text {N4532}}$ = 0.01, 0.1, 1.0.
\textbf{2nd row:} N(HI)$_{\text {N618}}$ = 0.7, 1.2, 5.0, 1.0, 2.0. N(HI)$_{\text {ESO}}$ = 0.1, 0.7, 5.0. N(HI)$_{\text {N4490}}$ = 0.7, 7.0, 70.0. 
\textbf{3rd row:} N(HI)$_{\text {N3448}}$ = 0.7, 1.2, 5.0, 1.0, 2.0. N(HI)$_{\text {U9560}}$ = 0.7, 3.0, 10.0.
\textbf{4th row:} N(HI)$_{\text {N672}}$ = 0.7, 1.2, 5.0, 1.0, 2.0. N(HI)$_{\text {N4449}}$ = 0.45, 1.0, 10.0. See \citet{putman03} for Galactic coordinates of the LMC/SMC and see Figure \ref{fig:IsoEnv} for deeper data of N4449 (here we show less deep data to include DDO125). For details on the HI observations and column densities: see Table \ref{table:HI}.}
\label{fig:sample}
\end{figure*}

\begin{figure}
\centerline{\includegraphics[width=\columnwidth]{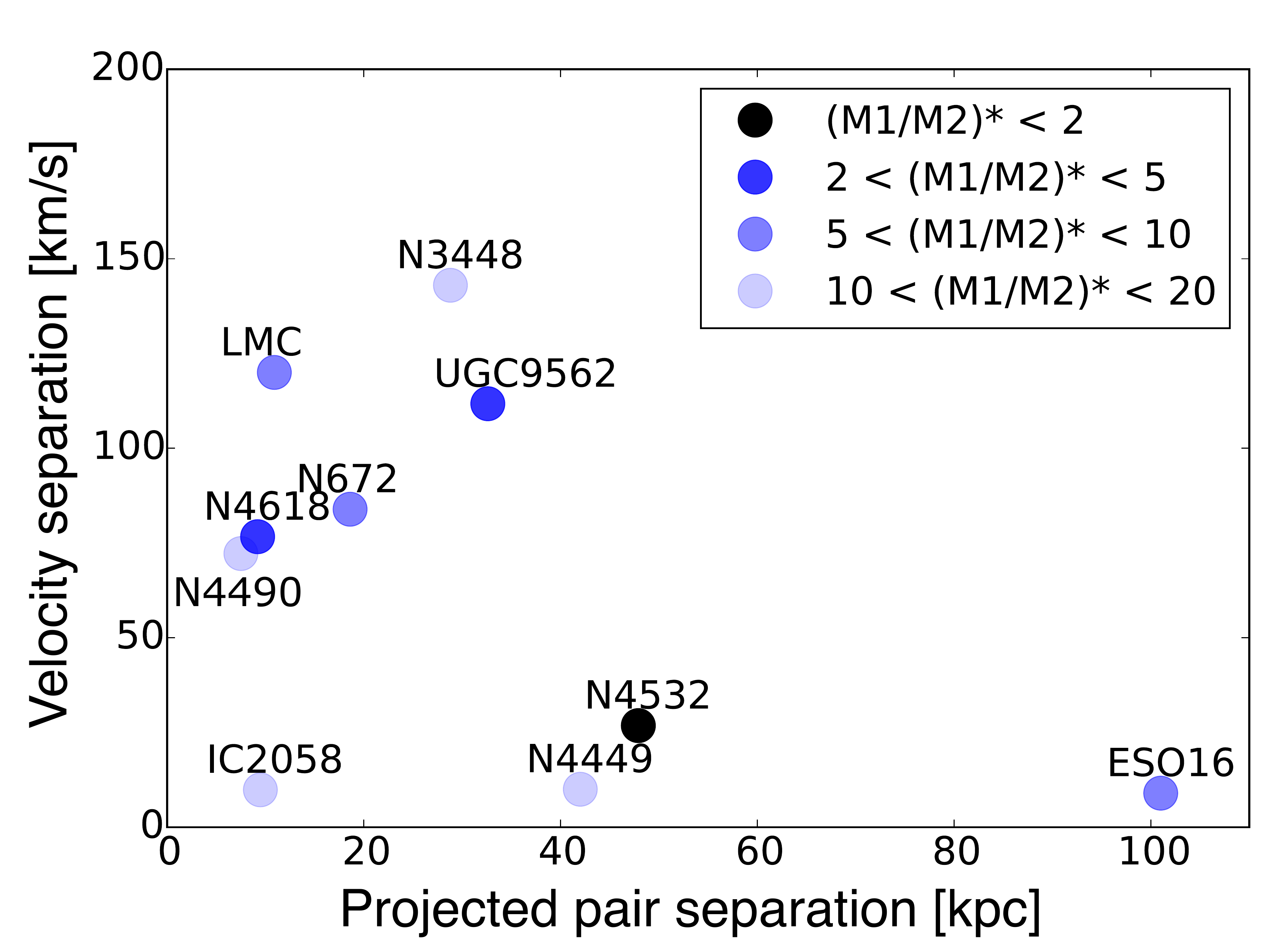}}
\caption{Projected radial and line of sight velocity separations for the 10 dwarf pairs colour coded by stellar mass ratios between the two dwarfs. Each data point is named by the primary (most massive) galaxy. All dwarf pairs are separated by less than 200 km s$^{-1}$ in their line of sight velocities and their radial separations are all smaller than 101 kpc.}
\label{fig:mass}
\end{figure}

Our main goals in this paper are to quantify the relative roles of environment and dwarf-dwarf interactions in removing gas to large radii (e.g. tidally or through ram pressure stripping), to understand how much material can be removed through these processes and determine whether this material remains bound to the pairs. We will do this through a systematic comparison of the HI content of 10 dwarf galaxy pairs utilizing the following terms:

\begin{itemize}
\item[-]{\textbf{Stellar extent of galaxies}}: as we are interested in disentangling the atomic gas in the outskirts of the dwarf galaxies from the gas residing within the galaxies, we use the 2MASS Extended Source Catalog to define the stellar extent of each galaxy. The extent is derived from the K$_s$-band scale length of each galaxy and is $\sim$ 4 $\times$ the K$_s$-band scale length. An elliptical fit is made by extrapolating the radial surface brightness from the standard isophote.\footnote{See \url{http://www.ipac.caltech.edu/2mass/releases/allsky/doc/sec4_5e.html} for details.} These fits are all illustrated as red ellipses on the HI profiles of the dwarf galaxies in Figure \ref{fig:sample}.\footnote{For the 2 of our 20 galaxies that did not have 2MASS K$_s$-band info (DDO137 and UGC 6016), we estimated the extent of the galaxies based on their r-band images. As the r-band traces younger stars, this might introduce a bias towards slightly larger radii.} This extent is used as the inner stellar component of each dwarf, and permits a uniform definition of the sizes of the main body of each target dwarf galaxy. 

\item[-]{\textbf{Distances}}: for each pair in our sample we use distances to the primary (most massive) dwarf from recent literature as listed in Table \ref{tab:local}. For three systems where there were no individual distance estimates (the ESO435-IG16 pair, the NGC 3448 pair and the UGC 9562 pair), we used the kinematic flow distance as listed in NED, which were corrected per \citet{mould00}, assuming $H_0$ $=$ 73 km/s/Mpc. In our tables we show how various parameters scale with distance. For the LMC and SMC, precise distance measurements exist for both galaxies, hence we use these individually for the two dwarfs. 

\item[-]{\textbf{Stellar masses}}: to determine the stellar masses of the dwarf galaxies, we used the K$_s$-band magnitudes from the 2MASS Extended Source Catalog of each galaxy and the conversion from light to mass defined in \citet{bell03} as:
\begin{eqnarray}
%\begin{equation}
(M/L)_{\odot}= 0.95 \pm 0.03
%\end{equation}
\end{eqnarray}
Both this and the definition of the stellar extent is biased towards older stars, but provide uniform values for our entire sample.

\item[-]{\textbf{Tidal indices, $\Theta$}}: our target pairs are located in very different environments (some are close to a more massive tertiary galaxy, and some pairs do not have a more massive galaxy in their vicinity). In order to determine what environment our dwarf pairs reside in, we calculated the tidal index, $\Theta$, used to quantify environment in the ANGST studies by \citet{weisz11} and defined in \citet{kara98} as:
\begin{eqnarray}\label{index}
\Theta = \text{log}_{10}\left(\frac{M_{*} [\times 10^{11} \text{M}_{\odot}]}{D_{\text{project}}^3[\text{Mpc}]}\right)
\end{eqnarray}
where $M_{*}$ is the stellar mass of the nearest massive galaxy and $D_{\text{project}}$ is the projected distance from the dwarf pair to the nearest massive galaxy. The nearest massive galaxy is referred to as the host galaxy. Here, $D_{\text{project}}$ is converted from an angular separation to a projected physical separation of the primary (most massive) dwarf in the dwarf pair to the host galaxy at the distance of the pair. We list all projected distances to the hosts in Table \ref{table:hosts}. Thus, a higher $\Theta$ indicates a stronger influence from a nearby neighbor. We do not use velocity separations of the pair and host when calculating the tidal index.

In this work, we search for a nearby massive galaxy within the NASA Extragalactic Database (NED) and define a massive galaxy as any galaxy with $M_*$ $>$ 10$^{10}$ M$_{\odot}$, and we require that the velocity separation between the massive galaxy and the pair is $v_{\text{sep}} <$ 500 km s$^{-1}$. If our pairs do not have a massive galaxy within $D_{\text{project}}$ $<$ 1500 kpc with $v_{\text{sep}} <$ 500 km s$^{-1}$, we classify the dwarf pair as being isolated and do not calculate the tidal index (see Table \ref{table:hosts} for details on the host galaxies). \citet{geha12} showed that dwarf galaxies in the field (D > 1.5 Mpc from a massive host) are not quenched, hence we apply this criteria for isolation. Similarly, studies for massive galaxy pairs (e.g. \citealt{patton13}) and for dwarf galaxy pairs (\citealt{stierwalt15}) also use this isolation criterion in addition to a velocity cut of $v_{\text{sep,host}} <$ 1000 km s$^{-1}$. As our sample is local, we use a stricter limit on $v_{\text{sep,host}}$ and our high tidal index pairs all have $v_{\text{sep,host}} <$ 300 km/s to their hosts (see Table \ref{table:hosts}). 

\item[-]{\textbf{Ram pressure stripping}}: hallmarks of ram pressure stripping are asymmetric extended structures (such as one sided trailing tails) and asymmetrically truncated disks (which is also seen for more massive galaxies as they fall into clusters,  \citealt{chung07}). To assess whether the dwarfs in our sample are experiencing ram pressure, we investigate if the HI surface density profiles in the inferred direction of motion deviate from the other directions (e.g. if the disks appear truncated). To quantify whether ram pressure can explain the features we see in the HI maps, we investigate the surface density profiles (see below) and use the simplified version of the \citet{gunn72} criterion as used by \citet{vollmer08} to investigate the halo densities necessary to explain any truncations:
\begin{eqnarray}\label{eq:voll}
\rho_{\text{IGM}} \sim \frac{v^2_{\text{rot}} \Sigma_{\text{gas}}}{v^2_{\text{gal}}R} \text{cm}^{-3}
\end{eqnarray} 
where $v_{\text{rot}}$ is the rotational velocity of the galaxy, $v_{\text{gal}}$ is the motion of the galaxy through the halo of the host, $\Sigma_{\text{gas}}$ is the column density of the gas at the truncation radius, and $R$ is the truncation radius. 

\item[-]{\textbf{Surface density profiles}}: in order to investigate the variations in the HI distributions of pairs in various environments, we compute the radial surface density distributions of the neutral gas for all dwarfs using the $MIRIAD$ ellint task. The task integrates the HI surface density maps in elliptical annulus, and we compute the average surface density in each annulus. To estimate how asymmetric the HI distributions are, we use various regions of the HI maps and integrate these separately (see Section \ref{sec:results}). 

\item[-]{\textbf{Dark matter masses from abundance matching}}: as we are interested in whether the pairs are bound to each other and whether the gas in the outskirts of the galaxies is bound to the pairs, we need an estimate of the total dark matter masses of our galaxies. For this purpose, we use the relation for abundance matching from \citet{moster13} using their Equation 2 and best fit values listed in their Table 1 at $z$ = 0. We assume a WMAP7 $\Lambda$CDM cosmology with ($\Omega_m$, $\Omega_{\Lambda}$, $\Omega_b$, $h$, $n$, $\sigma_s$) = (0.272, 0.728, 0.046, 0.704, 0.967, 0.810). Abundance matching assumes that the galaxy in question is isolated, which our galaxies are not. Furthermore, abundance matching is poorly calibrated for low mass galaxies (see e.g. \citealt{garrison14}). However, these estimates of the dark matter masses are only to be used as an approximation to the potential tidal fields of both the massive host and dwarf galaxies.

\item[-]{\textbf{Escape velocities}}: from the abundance matching we obtain an estimate of the total halo masses of our systems. That enables us to calculate the escape velocities from both the pairs (which can help quantify whether the gas in the outskirts of the pairs is bound) and from the primary, most massive dwarf (which enables us to estimate if the galaxies in the pair are bound to each other). We therefore adopt an NFW profile to the galaxies:

\begin{eqnarray}\label{NFW}
\Phi_\text{NFW}(R) = - \frac{GM_{200}ln(1 + \frac{R}{Rs})}{R f(c)}
\end{eqnarray}
which enables us to calculate the escape velocity at a specific distance from the primary dwarf:

\begin{eqnarray}\label{escp}
v_\text{escape} = \sqrt{2 \times \Phi_{\small\text{NFW}(R)}}
\end{eqnarray}
Here $R_s$ = $\frac{R_{200}}{c}$ is the scale length of the halo, where $R_{200}$ is defined as the radius at which the density is 200 times the critical density of the Universe, $M_{200}$ is calculated based on Eq. 2 in  \citet{moster13} using the cosmological parameters listed above and defines the mass within $R_{200}$,  $f(c) =  ln(1 + c) - c/(c+1)$ and the halo concentration, c, is defined from Eq. 4 in \citet{neto07}.

\item[-]{\textbf{HI bridges}}: bridges that connect two galaxies are strong indicators of an ongoing tidal interaction (e.g. \citealt{toomre72}, \citealt{combes78}, \citealt{hibbard95}, \citealt{barnes98}, \citealt{gao03}, \citealt{besla10}, \citealt{besla12}). In this work, we define a bridge as being continuous in HI column density and having a velocity gradient that smoothly connects one galaxy to the next (e.g. as seen in the Magellanic System, \citet{putman03}).

\end{itemize}

%---------------------------------------------------------------------Figure----------------------------------------------------------------

%---------------------------------------------------------------------Figure----------------------------------------------------------------

%---------------------------------------------------Results-----------------------------------------------------------------
\section{Results} \label{sec:results}
In this section we describe the results of our analyses of the diffuse gas in our LV-TNT sample. First we examine the observational differences between the neutral gas distributions of the dwarf galaxy pairs evolving in isolation and in the vicinity of massive hosts (see Section \ref{sec:obs}). In Section \ref{sec:global} we examine the global trends of gas removal for our entire sample.

%---------------------------------------------------------------------Figure----------------------------------------------------------------
\begin{figure*}
\centerline{\includegraphics[width=\textwidth]{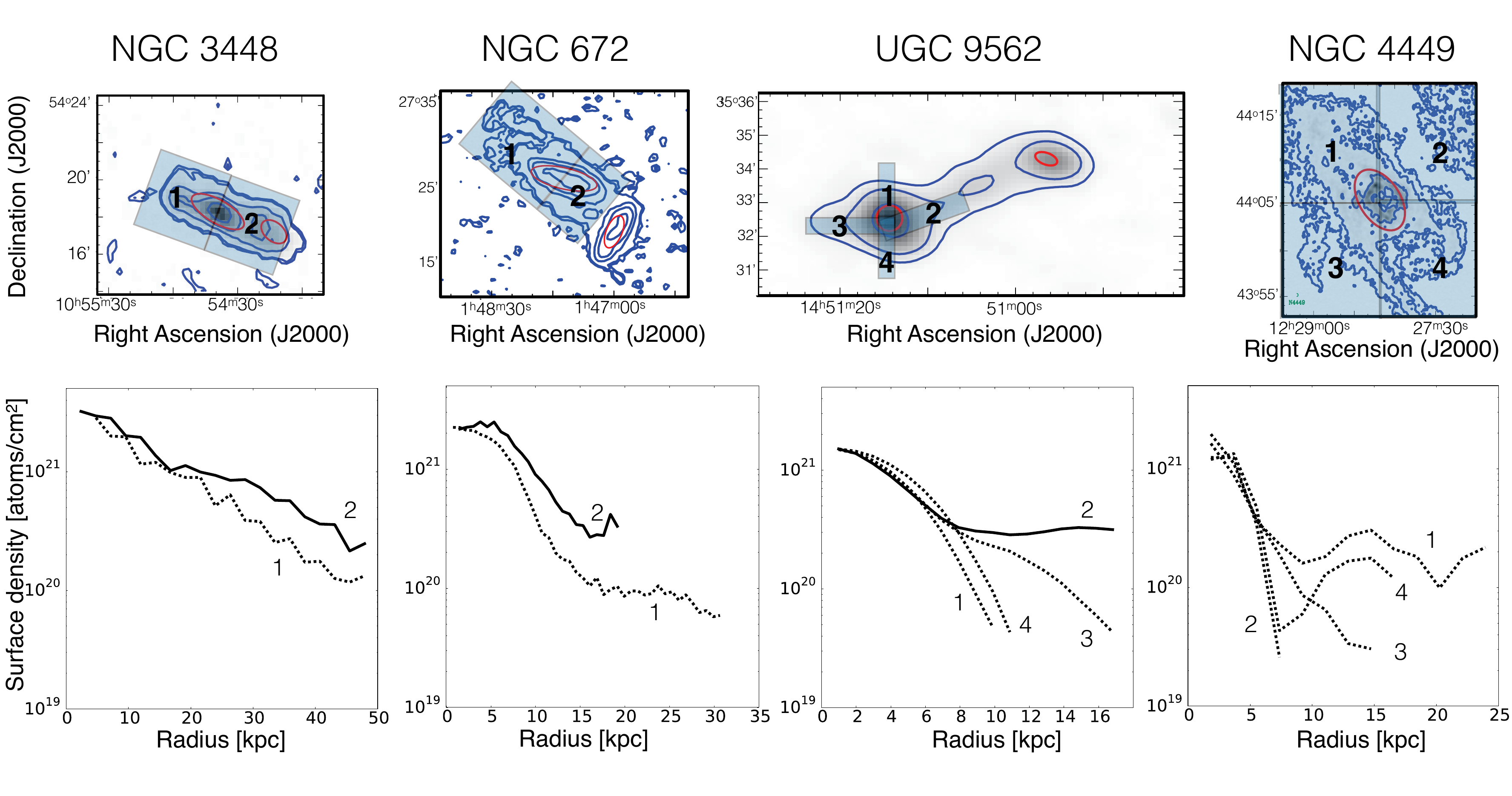}}
\caption{\textbf{Top row:} HI contours (blue) of the dwarf pairs in our sample that are isolated with $\Theta <$ 0. The 2MASS extents of all dwarfs are shown in red ellipses. Here we used deeper data from \citet{hunter98}, zoomed in on N4449. The numerated boxes indicate regions of the primary dwarf for which we compute HI surface density profiles (see below).
\textbf{Bottom row:} HI surface density profiles vs radial distance from 2MASS ellipse centre in two or four different directions (see numerated boxes) for each primary dwarf galaxy in each pair. We use two lines if the inclination of the primary dwarfs are high. Solid lines indicate the direction of the bridge connecting the primary to the smaller companion. All profiles show a flattening towards the smaller companion (due to the higher densities in the bridges), however their HI distributions do not show indications of material stripped by ram pressure (no rapid drops to lower column densities in profiles and no asymmetric trailing features). Since there is no bridge connecting NGC 4449 and DDO125, no solid line is shown on the plot.}
\label{fig:IsoEnv}
\end{figure*}
%---------------------------------------------------------------------Figure----------------------------------------------------------------

\subsection{Distribution of Neutral Gas as a Function of Environment }\label{sec:obs}
To understand how dwarf-dwarf interactions are affected by their environment, we first examine the influence of dwarf-dwarf encounters on the extended gas structure in Section \ref{sec:iso} using isolated dwarf pairs. Subsequently, we search for observational evidence of environmental processing (ram pressure stripping) in the extended gas structure of the non-isolated dwarf pairs in Section \ref{sec:inter} and \ref{sec:dens}. We investigate how these non-isolated dwarf pairs differ from dwarf pairs evolving in isolation, and we break down this analysis in terms of tidal indices. The tidal index is an estimate of the tidal influence of a massive host galaxy on the dwarf galaxy pairs. Here, we define isolated pairs as having $ \Theta < 0$, intermediate tidal index pairs as having $ 0 < \Theta < 1.5$ and high tidal index pairs as having $\Theta > 1.5$ (see Table \ref{table:hosts} for details on each massive host galaxy). We checked that the only pair potentially affected by the distance errors, which could move it to a lower tidal index group, was the NGC 4532 pair (see Appendix~\ref{sec:N4532}).

\subsubsection{Dwarf pairs evolving in isolation}\label{sec:iso}
By investigating the dwarf pairs evolving in isolation, far from a massive host, we can understand how the interaction between the dwarfs affects the gas distributions. In Figure \ref{fig:IsoEnv} (top row) we show the four pairs in our sample with $\Theta < 0$: NGC 3448 \& UGC 6016, NGC 672 \& IC 1727,  UGC9560 \& UGC 9562 and NGC 4449 \& DDO 125 (from left to right). 

All isolated pairs, except for the NGC 4449 pair, show gaseous bridges connecting the two dwarfs indicating a mutual tidal interaction (e.g. \citealt{toomre72}, \citealt{combes78}, \citealt{hibbard95}, \citealt{barnes98}, \citealt{gao03}, \citealt{besla10}, \citealt{besla12}). Interestingly, NGC 4449 and its companion DDO 125 have the largest separation of the isolated pairs, which could explain why a bridge is not present at this interaction stage. The bridges in the three other pairs are all continuous in HI column density and have smooth velocity gradients across the bridges. The surface density profiles of the first three primary isolated dwarfs (Figure \ref{fig:IsoEnv}: lower row) flatten out towards the smaller companion (where dense bridge material is present). The existence of bridges in these isolated pairs implies that material is being removed from the lower mass companion due to the presence of the nearby dwarf (i.e. gas is being pre-processed) and that this process takes place without the aid of other environmental factors (such as a nearby massive host galaxy). While the NGC 4449 pair does not have a bridge, the HI map shows signs of interaction through its gaseous arm surrounding its stellar component (see Appendix~\ref{sec:N4449} for more details). 

Thus, based on our limited sample dwarf-dwarf interactions alone do not seem to create asymmetric extended tails or obvious truncation of the gaseous disks, but tidal pre-processing appears to be taking place.

\subsubsection{Intermediate tidal index pairs}\label{sec:inter}
The dwarfs evolving in the vicinity of massive host galaxies can give insight to how the environmental effects will influence the dwarf pair interactions. We therefore search for evidence of asymmetries indicating ram pressure stripping (such as truncated HI disks and one sided trailing tails) in the non-isolated pairs.

\begin{figure*}
\centerline{\includegraphics[width=\textwidth]{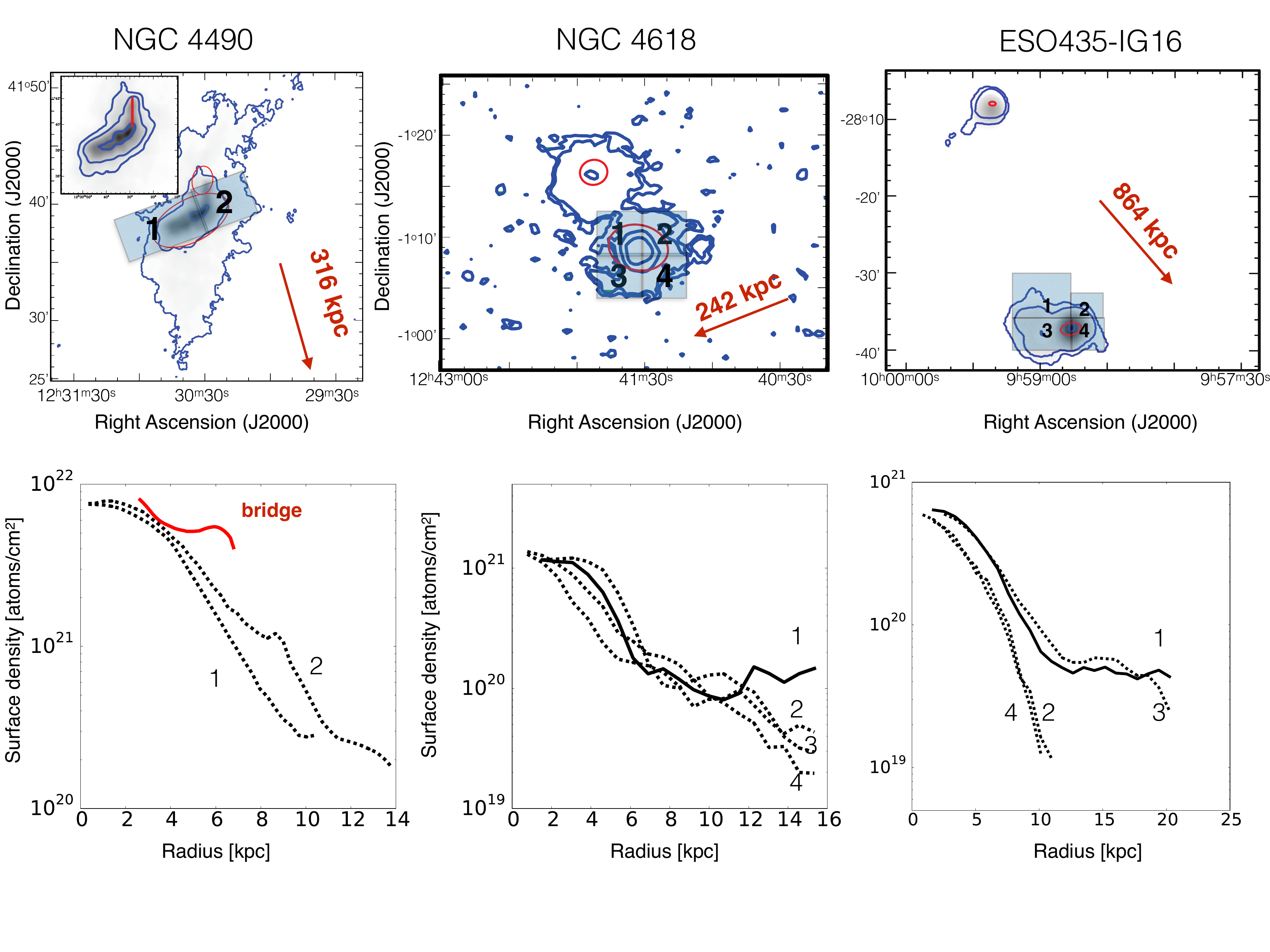}}
\caption{\textbf{Top row:} HI contours (blue) of the dwarf pairs in our sample that have $0 < \Theta <$ 1.5. The 2MASS extents of the dwarfs are shown in red ellipses and the location of the massive host is illustrated by red arrows along with the projected distance to the host in kpc. The numerated boxes indicate regions of the primary dwarf for which we compute HI surface density profiles (see below).
\textbf{Bottom row:} HI surface density profiles vs radial distance from the 2MASS ellipse centre in different directions (see numerated boxes) for each primary (most massive) dwarf galaxy in each pair. Solid lines indicate the direction of the bridge connecting the primary to the smaller companion. Due to the high inclination of the NGC 4490 system (left), we plotted the surface density using only two regions, and plot the density in the bridge separately (see zoomed box and red solid line). The surface density profiles show a flattening towards the smaller companion, however their HI distributions do not show indications of material stripped by ram pressure (no rapid drops to lower column densities in the HI profiles in the direction of the massive hosts, as compared to towards the other directions and no asymmetric trailing features in the envelopes nor surface density profiles).}
\label{fig:InterEnv}
\end{figure*}

In Figure  \ref{fig:InterEnv} we show the HI maps of the 3 pairs in our sample with $0 < \Theta < 1.5$: NGC 4490  \& NGC 4485, NGC 4618 \& NGC 4625 and ESO435-IG16 \& ESO435-IG20. The red arrows indicate the direction of the nearest massive host, which the pairs could in principle be moving towards. The HI distributions of NGC 4490  \& NGC 4485 and NGC 4618 \& NGC 4625 are both symmetric, with no hint of an extended one-sided tail or a truncation in any direction. NGC 4618 \& NGC 4625 have overlapping HI disks, which is also seen in the flattening of the surface brightness profile of NGC 4618 in this direction (Figure \ref{fig:InterEnv}: bottom, right panel, region 1). However, it is unclear from the velocity gradient in the WHISP maps, whether the gas connecting the two galaxies is a bridge or just a projection effect of overlapping HI profiles, as there is only a vague indication of a smooth gradient (\citealt{hulst01}).

For NGC 4490 \& NGC 4485 a very extended, low density envelope surrounds the pair. Given the symmetric nature of the envelope it is unlikely that it is present due to ram pressure stripping. The dense bridge connecting the two dwarfs indicates an ongoing tidal interaction, which could explain the extended envelope (see red, solid line Figure \ref{fig:InterEnv}). We do not see envelopes in the other isolated and intermediate tidal index pairs, but those have larger pair separations. Thus, it is not surprising to find the strongest tidal signatures for the NGC 4490 pair, as the dwarfs in this pair are located very close to each other.

In contrast to the other pairs, the ESO435-IG16 pair does not have a bridge connecting the two galaxies, however there is extended material from the primary in the direction of the secondary and from the secondary in the direction of the primary, which suggests a tidal interaction between the two galaxies. The ESO435-IG16 pair is a widely separated system ($\sim$ 100 kpc),  so it is possible that a dense bridge existed at some point, but now has too low of a column density (since stretched over such a large distance). The surface density profiles in region 2 and 4 of ESO435-IG16 deviate from the other two directions (they drop to lower column densities at smaller radii), which could indicate a truncation in this direction (this is the same direction as the galaxy NGC 3056, which is $\sim$ 220 kpc away). However, as the pair is located $>$ 2 $\times$ R$_{200}$ (the radius at which the density of the host is 200 $\times$ the critical density of the Universe) of NGC 3056, it is unlikely that ESO435-IG16 is truncated due to ram pressure stripping (see Section \ref{sec:cgm} for an analysis of this).

Hence, the three pairs with $0 < \Theta < 1.5$, do not appear to be affected by the closest massive host galaxy and resemble the pairs evolving in isolation,  i.e. that significant gas evolution is seen in the formation of bridges and extended envelopes, but there are no large scale asymmetries.

%---------------------------------------------------------------------Figure----------------------------------------------------------------
%%---------------------------------------------------------------------Figure----------------------------------------------------------------

\subsubsection{High tidal index dwarf pairs}\label{sec:dens}
\begin{figure*}
\centerline{\includegraphics[width=\textwidth]{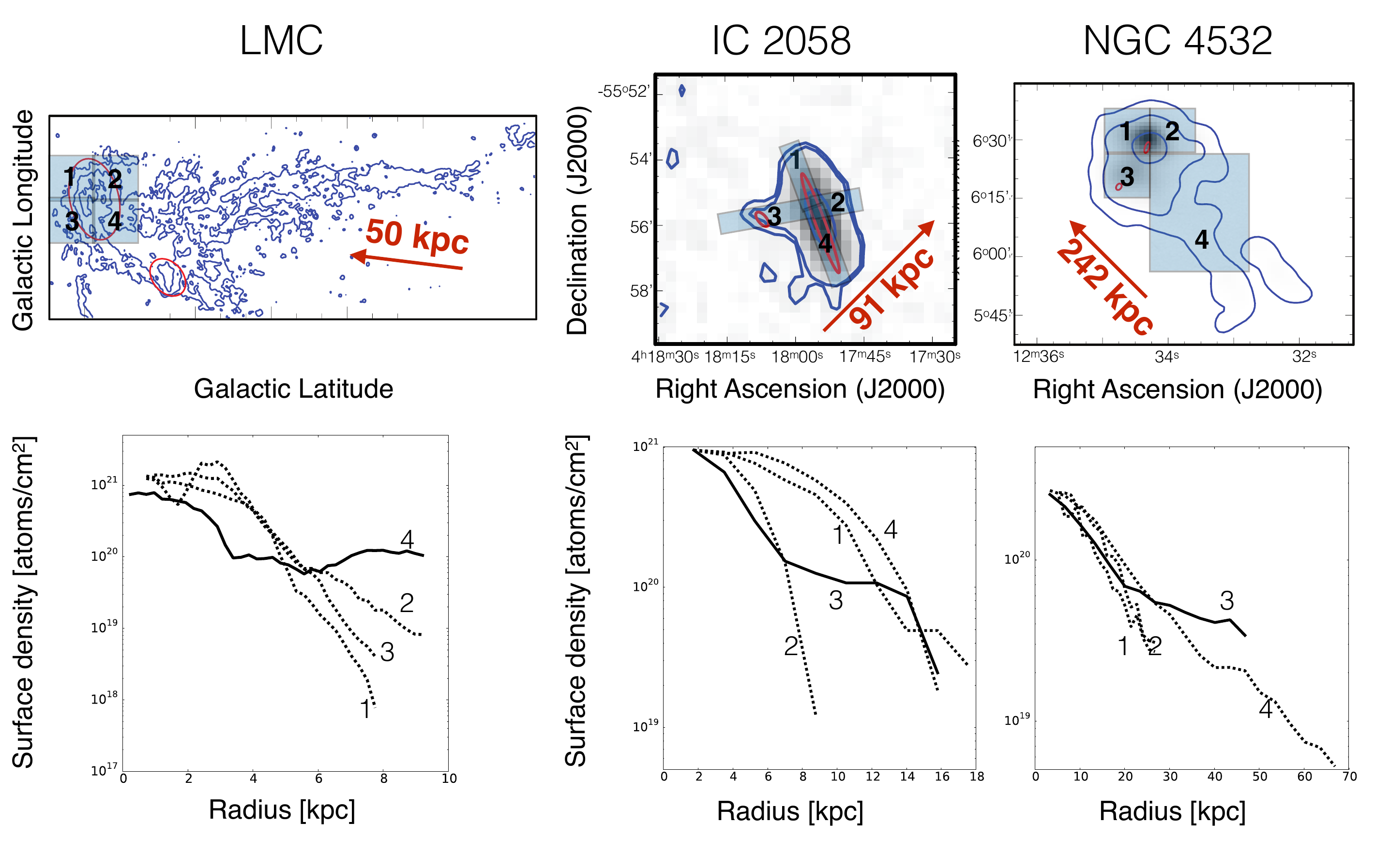}}
\caption{\textbf{Top row:} HI contours (blue) of the four dwarf pairs in our sample that have $\Theta >$ 1.5 indicated by the name of the primary dwarf. The 2MASS extents of all dwarfs are shown in red ellipses, and the assumed direction of motion (true for the LMC/SMC motion) through the halo of the massive host is illustrated by red arrows along with the projected distance to the host in kpc. The numerated boxes indicate regions of the primary dwarf for which we compute HI surface density profiles (see below). \textbf{Bottom row:} HI surface density profiles of the primary dwarfs (the most massive dwarf in pair) vs radial distance from the 2MASS ellipse centre in four different directions (see numerated boxes) for each primary dwarf galaxy in each pair. Solid lines indicate the direction of the bridge connecting the primary to the smaller companion. All profiles except for NGC 4449 show a flattening towards the smaller companion (due to the higher densities in the bridges), all profiles show rapid drops in column density in the direction towards the massive host galaxy and all profiles are extended in the opposite direction of the host (due to trailing tails, except for IC 2058 for which the companion is located in this direction making it difficult to disentangle the effect of the companion vs trailing material).}
\label{fig:DenseEnv}
\end{figure*}

The dwarf galaxy pairs that have tidal indices, $\Theta >$ 1.5, indicating that they are not isolated from the influence of a massive neighbor are LMC \& SMC, IC 2058 \& PGC 75125, NGC4352 \& DDO137 (see Figure \ref{fig:DenseEnv}). These are the most likely pairs to be affected by ram pressure stripping and tides from the hosts in our sample.

In Figure \ref{fig:DenseEnv} we show the HI maps of the three pairs with $\Theta > 1.5$ (see Table \ref{table:hosts} for details on each massive host galaxy). These pairs are all within R$_{200}$ of their hosts. The red arrows illustrate the inferred direction of motion of the dwarf pairs (in the case of the LMC we know this from HST proper motion observations (\citealt{nitya09}), and for the other two pairs, we assume that the dwarfs are moving towards their host galaxy).
%---------------------------------------------------------------------Figure----------------------------------------------------------------
%---------------------------------------------------------------------Figure----------------------------------------------------------------

From Figure \ref{fig:DenseEnv} (top row) it is evident that the morphologies of the HI distributions all show asymmetries and have material stripped preferentially in the opposite direction of the inferred direction of motion. The LMC and NGC 4352 systems in  particular have diffuse tails that extend in the opposite direction of their direction of motion (inferred direction for the NGC 4532 pair). For the LMC pair the diffuse tail is detectable at column densities $>$ $7 \times 10^{19}$ atoms cm$^{-2}$, where the diffuse tail is only detected at column densities $<$ $7 \times 10^{18}$ atoms cm$^{-2}$ for the NGC 4532 pair. If the dwarf galaxy pairs are in fact moving through the halo material of the more massive host galaxies (as we know is the case for the LMC \& SMC) this extended material has the shape that we would expect from ram pressure stripping (trailing material), which is also observed for the HI distributions of massive spiral galaxies as they fall into clusters (e.g. \citealt{chung07}). For the LMC \& SMC, \citet{putman98} found a low surface density leading neutral gas component to the Magellanic Stream indicating that the tidal forces along with ram pressure stripping is shaping the MS. Whether leading tidal streams are also present in our other high tidal index pairs is not clear from the level of sensitivity of these HI maps. 

In Figure \ref{fig:DenseEnv} (bottom row) we demonstrate how the average HI surface density profiles of the primary dwarf galaxies vary in different directions (see numerated regions centred on the 2MASS ellipses of each primary dwarf galaxy). Interestingly, the surface density profiles all flatten out in the direction towards the smaller companion dwarf where a bridge is present (solid lines). This was also seen for the isolated and intermediate tidal index pairs. 

The HI surface density profiles all show a feature similar to a truncation in the inferred direction of motion.
This could indicate that ram pressure stripping of the dwarf pairs by halo material from the host might be occurring. While the HI distributions could be extended and asymmetric due to mutual interactions between the pairs alone, truncated HI disks are difficult to explain without ongoing ram pressure stripping, and we explore the possibility of ram pressure stripping in Section \ref{sec:cgm}. Interestingly, no features of truncated disks were found in the isolated or intermediate tidal index pairs.

Lastly, the surface densities in the direction opposite of the inferred direction of motion are extended (we see trailing material). In IC 2058 this is difficult to assess due to the coincident location of the secondary dwarf galaxy and the high inclination of IC 2058. 

Thus, there are asymmetries in the HI distributions surrounding the dwarf pairs in our sample that are near massive host galaxies. Mutual tidal interaction between the dwarfs might have removed the majority of the HI gas and formed bridges in all examples, but the deviation (drop to lower column densities at smaller radii) in the HI surface density profiles in the direction of the host galaxy and the lack of detectable leading streams, indicate that ram-pressure stripping could play an important role in shaping the systems, which was not the case for the isolated and intermediate tidal index dwarf pairs.

\subsection{Global Trends: HI Extent and Gas Mass}\label{sec:global}
To better understand the global gas removal processes and to quantify the features we found in the previous sections, we here investigate the amount of gas outside each pair, and its extent compared to non-paired analogs.

\subsubsection{Quantifying the amount of gas outside the dwarf pairs}\label{sec:outer}
\begin{figure}
\centerline{\includegraphics[width=\columnwidth]{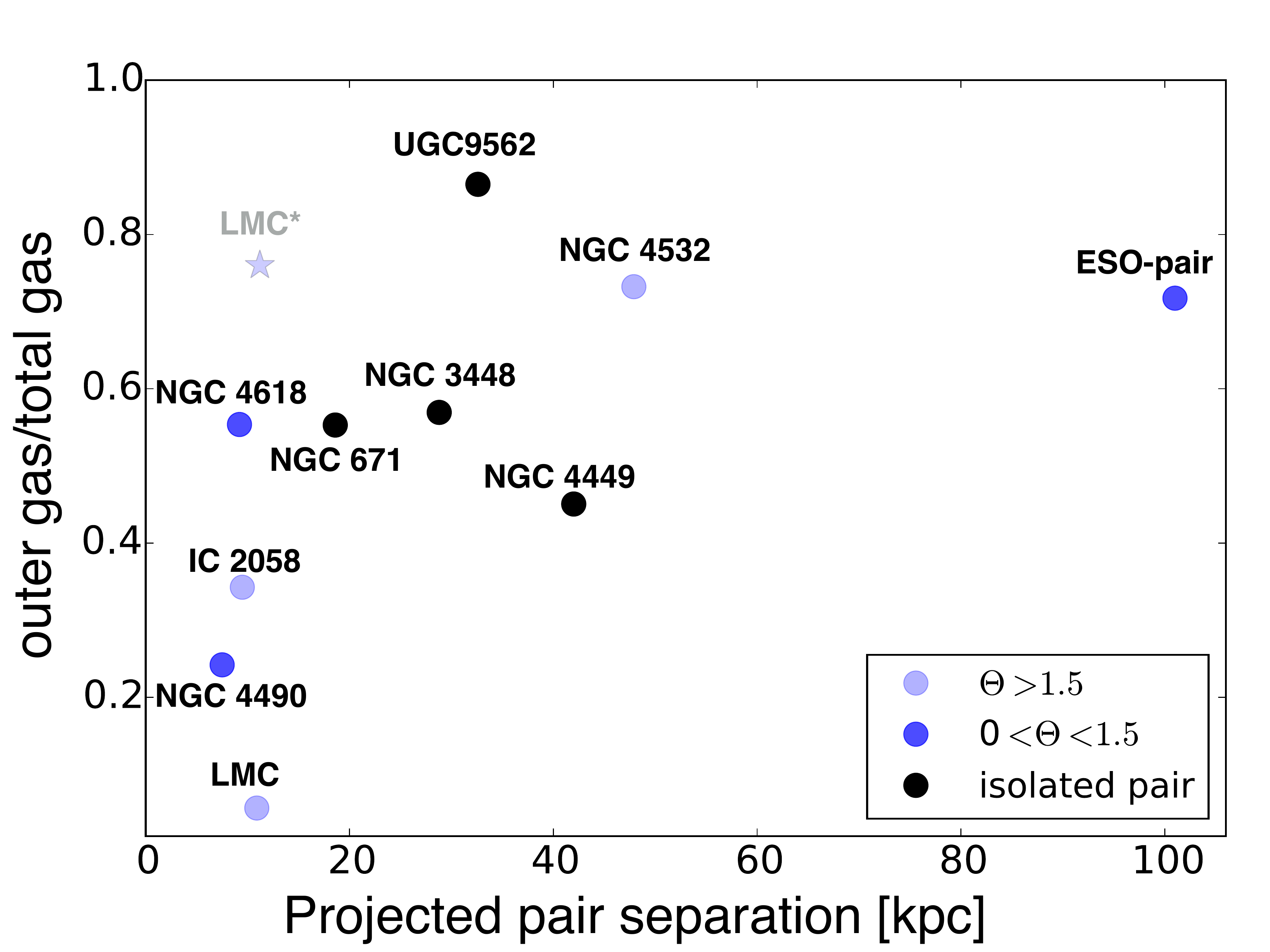}}
\caption{Fraction of neutral gas outside the 2MASS extents of all dwarf pairs to their total HI mass for each pair system, colour coded by their environment (tidal index, $\Theta$). The data points are labeled based on the name of the primary dwarf in the pairs. A uniform column density cut of N(HI) = 7 $\times$ 10$^{19}$ atoms cm$^{-2}$ was made on all maps to ensure a systematic comparison. The two pairs close to a massive host (light blue) have a small fraction of gas residing outside their stellar disks. The star indicates the fraction of HI and ionized gas (\citealt{fox14}) outside the 2MASS extent of the LMC/SMC pair to the total HI gas in the pair and the ionized gas outside the pair.}
\label{fig:outer}
\end{figure}

In Figure \ref{fig:outer} we show the amount of gas residing outside the 2MASS ellipses of our dwarf galaxies in each pair compared to the total amount of gas in the dwarf pair system (the data points are named by the primary dwarf in each pair, but represent the amount of gas outside both pair members). In addition to making a uniform definition of the inner vs outer part of the galaxy for all dwarfs (the 2MASS ellipses), we also define a uniform column density cut to each data set (N(HI) = $7 \times 10^{19}$ atoms cm$^{-2}$) to enable a systematic comparison between all pairs. 

We find that the two highest tidal index pairs (LMC \& SMC and IC 2058 \& PGC 75125), which have twice the tidal index value of the other dense environment dwarf pairs and are the two closest to their hosts, have low neutral gas mass fraction of outer vs total gas. The other dwarf galaxy pair that showed signs of ram pressure (the NGC 4532 pair) has a large amount of gas in its outskirts compared to the total system, which is also the case for the more isolated galaxies. The NGC 4532 pair is farther away from its host than LMC/SMC and its host galaxy is less massive than the Milky Way, as such, it is possible that most of the gas removed from the LMC and IC 2058 pairs is in an ionized state, whereas the gas associated with NGC 4532 is neutral. Interestingly, if we were to add the amount of ionized gas found in the MS (\citealt{fox14}) to the total amount of gas in the MS (ignoring the uniform cut in column density for the ionized gas), the outer/total gas of the LMC would be $\sim$ 0.76 (see star in Figure \ref{fig:outer}), which is very similar to the NGC 4532 pair (which has outer/total = 0.74). This supports the idea that the HI in the NGC 4532 tail has not yet been ionized (see HST-COS result presented in Appendix~\ref{sec:N4532}). 

While the majority of the isolated and intermediate tidal index pairs have gas mass fractions of their outer vs total neutral gas $>$ 50\%, the NGC 4490 \& NGC 4485 pair has a low amount of outer to total gas despite the large diffuse envelope surrounding the pair. Due to the small separation between the two dwarf galaxies, it is important to note that the bridge material connecting them is included within the 2MASS ellipses of the dwarfs leading to a higher amount of inner neutral gas in our estimates. As mentioned earlier the envelope is of low column density and the amount of mass in it is not substantial compared the amount of neutral gas residing in the dwarf galaxies ($\sim 30 \%$ of the neutral gas is in the envelope). This could indicate that the tidal interaction between the two galaxies did not actually remove a substantial amount of gas, that the gas has been tidally removed but has started to fall back in, or that a substantial amount of the removed gas has been ionized. To differentiate between these processes, detailed modeling of the dynamics and gas physics is necessary (Pearson et al., in prep).

%%---------------------------------------------------------------------Figure----------------------------------------------------------------
%%---------------------------------------------------------------------Figure----------------------------------------------------------------

\subsubsection{The extent of the neutral gas compared to non-paired analogs}\label{sec:extent1}
As a control sample, we use the late type dwarf irregulars from the \citet{swaters02} sample (hereafter S02). This sample consists of 73 galaxies that were selected from the Uppsala General Catalogue of Galaxies, requiring absolute magnitudes fainter than $M_B$ = 17, which had HI flux densities larger than 200 mJy and galactic latitudes $>10^\text{o}$. These 73 galaxies were not selected based on any environmental criteria, hence some of them could be close to massive host galaxies. Additionally, a small fraction of the dwarfs in the S02 sample are in dwarf pairs (e.g. IC 1727 and NGC 4449 which are also in our sample). To understand how the HI extents of our dwarf irregulars are affected by their mutual interaction, we investigate how they compare globally to the S02 sample.

\begin{figure}
\centerline{\includegraphics[width=\columnwidth]{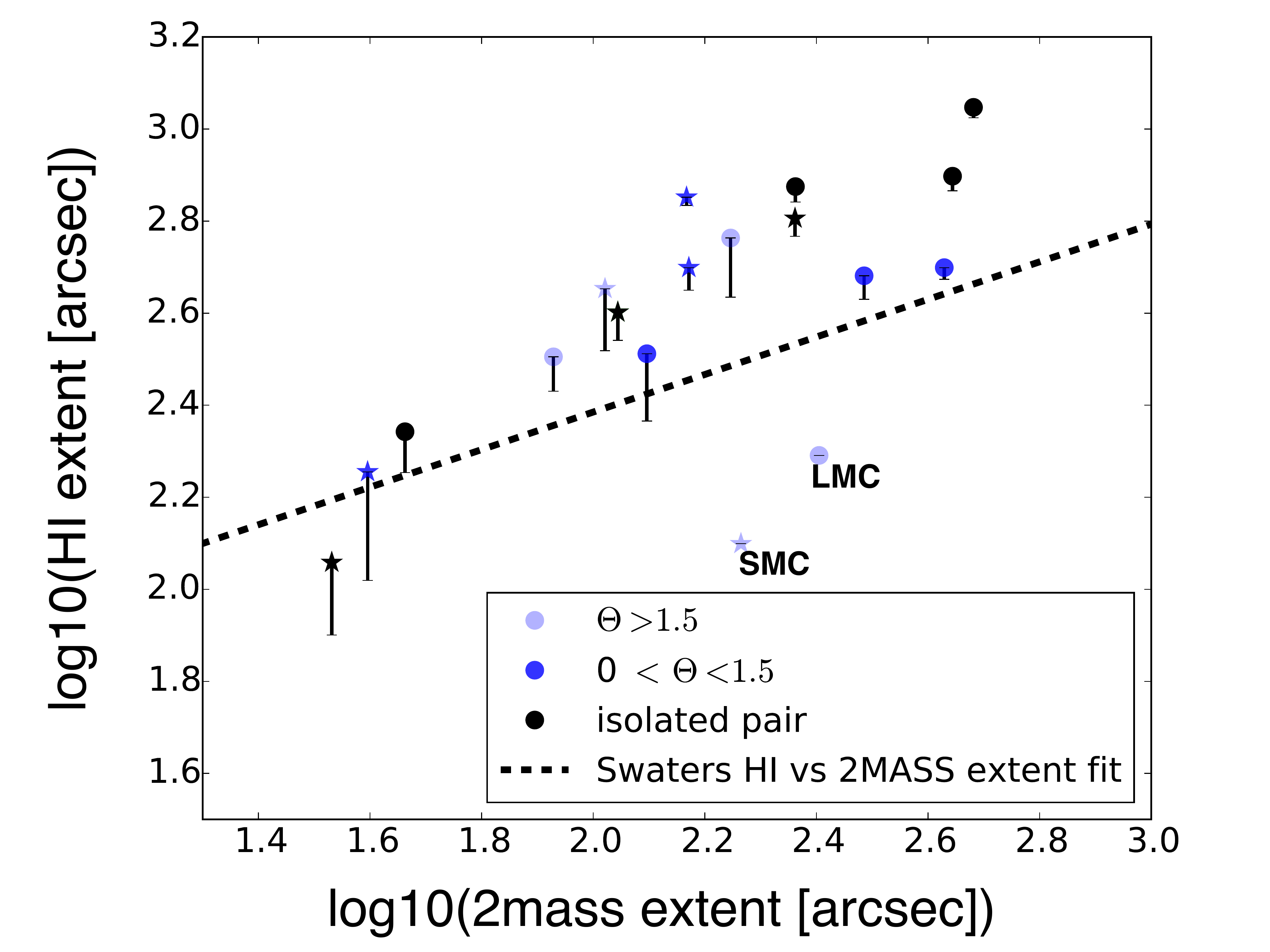}}
\caption{The HI extent of all the dwarf galaxies at N(HI) = 1.2 $\times$ 10$^{20}$ atoms cm$^{-2}$ plotted vs the 2MASS extent of all the dwarf galaxies in our sample colour coded by environment (tidal index, $\Theta$). The circles represent the primary dwarf galaxies and the stars represent the secondary dwarf galaxies. 
 The dashed line shows the HI diameter vs the 2MASS stellar extent fit for the 20 non-paired dwarf galaxies in the \citet{swaters02} sample that have defined 2MASS stellar extents. To avoid extrapolating the \citet{swaters02} fit, the LMC/SMC have been scaled to a distance of 11 Mpc. The error bars indicate the the range of HI extents possible due to potential beam dilution. We find that most of our dwarfs fall above the fit indicating that the high column density gas is tidally extended.}
\label{fig:extent}
\end{figure}

Figure \ref{fig:extent} shows the DHI extent (defined as the diameter at which N(HI) = 1.2 $\times 10^{20}$ atoms cm$^{-2}$)  of our galaxies vs their stellar extents (defined as the extent from the 2MASS extended source catalog). These values are also listed in Table \ref{table:prop}. Primary dwarf galaxies are marked by circles and secondary dwarf galaxies are marked by stars. The secondary dwarfs, DOO 125 and PGC 75125 were not included in this plot as DDO 125 is at the edge of the HI map and PGC 75125 is much smaller than the beam size (see Figure \ref{fig:sample}), which in both cases makes an assessment of the surface density profiles difficult. We note that due to the presence of the dense bridges, azimuthally smoothing of surface density profiles would lead to an overestimation of the DHI extent. Hence, here we use the DHI from the surface density profiles yielding the maximum extent DHI that is not in the direction of the bridge. If no apparent bridge is present, we use the azimuthally smoothed surface density profiles (i.e. for NGC 4449 and the ESO pair). For highly inclined systems (NGC 3448, UGC 6016, NGC 672, IC 1727 and IC 2058), we use the major axes for which bridge material is not present.

Of the 73 dwarf irregular galaxies in the S02 sample, 22 of them have stellar extents defined in the 2MASS extended source catalog, which enables us to compare them to our data. Of these 22 galaxies, two of them are in our dwarf pair sample (IC 1727 and NGC 4449). The dashed line in Figure \ref{fig:extent} is the HI diameter vs the 2MASS stellar extent fit for the 20 non-paired dwarf galaxies in the S02 sample that have defined 2MASS stellar extents. The log(2MASS extent) and the log(DHI extent) in the 20 non-paired dwarf galaxies range from 1.5 - 2.7 [arcsec] and 2.1 - 3.1 [arcsec] respectively. Instead of extrapolating the fit from the 20 non-paired dwarf galaxies to LMC/SMC angular scales of $\sim$ 30000 arcsec, we plot the LMC/SMC 2MASS vs HI extents scaled to a distance of 11 Mpc. The errors on the HI extents are not published for the S02 sample nor for the 2MASS extents for the dwarfs. Our HI extent estimates only differ from those in the S02 sample by having non-uniform beam sizes (see Table \ref{table:HI}), while the S02 beam sizes are all 60". Our HI extents could therefore be over-estimated from the emission not filling the entire beam, and we include this potential overestimate as errorbars on Figure \ref{fig:extent}. We do not include the equivalent for the S02 sample, but this would also cause the dashed line to have an uncertainty downward. Only three of our pairs have beam sizes larger than S02.

We find that the majority of our paired dwarf galaxies fall above the S02 fit (see Figure \ref{fig:extent}) and thus have more extended dense HI disks compared to their stellar extents than their non-paired counterparts. This is true even without having accounted for the HI bridge, which would only increase the DHI value. While extended HI disks are common for dwarf irregular galaxies (seen in large scatter found by S02), the fact that the majority of our dwarf galaxies systematically fall above the fit, suggests that tidal interactions between dwarfs moves gas to the outskirts of the galaxies. Hence, even the high column density neutral gas in these systems is ``loosened up'' (pre-processed) from tidal interactions, which might affect the efficiency of gas lost to the CGM of more massive galaxies if captured (see e.g. \citealt{besla12}, \citealt{salem15}). 

Interestingly, the LMC and SMC fall below the S02 fit, indicating that their gas disks have been truncated and that gas has been pushed to the central regions. As the LMC and SMC pair is the one closest to a massive galaxy in our sample, this provides interesting insight to the importance of pre-processing and how dwarf galaxy pairs eventually feed the massive galaxies they are falling into. 
Additionally, UGC 9560 (the star with the smallest 2MASS extent) falls slightly below the fit, however the beam is larger than the inferred DHI, which could be affecting our accuracy in determining the HI extent. Higher resolution data would help resolve this. 

None of the S02 dwarf irregulars show extended, one sided, trailing features to a depth of $\sim$ 7 $\times 10^{19}$ atoms cm$^{-2}$ (which are seen in the LMC/SMC and NGC 4532/DDO137 pairs), even though the S02 sample was not selected based on any environmental criteria. This further strengthens the argument that pre-processing via dwarf-dwarf tidal interactions are key in removal of gas to the outskirts, and that environmental processes can subsequently shape this extended material, causing truncations and dominant trailing streams.

%%---------------------------------------------------------------------Figure----------------------------------------------------------------
%%---------------------------------------------------------------------Figure----------------------------------------------------------------

%---------------------------------------------------Discussion-----------------------------------------------------------------
\section{Discussion}\label{sec:discussion}
In the previous section, we found that gas was moved to the outskirts of dwarf galaxies in tidal interactions (see Figure \ref{fig:outer} and \ref{fig:extent}) and through environmental effects (as seen in the extended tails present in the high tidal index pairs presented in Figure \ref{fig:DenseEnv}). Motivated by these results, we explore in Section \ref{sec:bound} whether this gas is unbound and fully removed from our dwarf pairs. Since ram pressure appears to play an important role in moving gas to the trailing tails in our high tidal index pairs, we investigate the host halo densities needed to produce these ram pressure signatures in Section \ref{sec:cgm}. Independent of the environment, the majority of the dwarf pairs have dense bridges connecting the individual galaxies. In Section \ref{sec:bridges} we therefore discuss the properties and formation mechanisms of these bridges. In Section \ref{sec:sfr} we explore the star formation rates in our sample, as outflows from star formation is a different channel of moving gas to the outskirts of galaxies. 

\subsection{Is the Gas Unbound in the Interactions?}\label{sec:bound}
The mechanisms  responsible  for  the morphological transitions of galaxies (e.g. the formation of dwarfs spheroidal galaxies) is still an open question. \citet{grc09} found evidence that ram pressure stripping is responsible for the transition of Local Group dwarf galaxies from gas rich to gas poor systems.  Others argue that dwarf spheroidals can be produced in mergers between disky dwarf galaxies (e.g. \citealt{ebrov15},  \citealt{kaz11}). 
Additionally, it has been suggested that the origin of a large amount of metals in the IGM (\citealt{danforth08}) could be due to outflows from the shallower potential wells of dwarf galaxies (e.g. \citealt{martin04}, \citealt{martin05}, \citealt{tolstoy09}). It is uncertain whether mutual tidal interactions between dwarfs can serve as a mechanism for gas removal (although the existence of actively star-forming dwarfs in the field  (\citealt{geha12}) suggests this is unlikely to remove all gas from the dwarfs). Removing gas through dwarf-dwarf interactions would be a different channel of transforming from gas rich dwarf irregulars to gas poor systems (e.g. dwarf spheroidals) and simultaneously feed the IGM with baryons. For more massive galaxies, \citet{barnes16} recently showed that the extended tails formed in tidal interactions remain bound to the merging galaxies. Using numerical simulations, \citet{bekki08} showed that dwarf-dwarf merging can trigger central starbursts and transform the merging dwarfs into blue compact dwarfs (BCDs). While the starburst will consume some of the gas, the newly formed BCDs were surrounded by massive extended HI envelopes, indicating that a substantial amount of HI mass remains bound and is not lost to the IGM in dwarf interactions. 

To explore the question of whether dwarf-dwarf interactions prior to final coalescence can facilitate the morphological transformation of gas rich irregular galaxies to gas poor dwarf spheroidals and whether their interactions feed baryons to the IGM, we utilize available velocity maps to assess whether the extended gas distributions surrounding our dwarf pairs are in fact bound to the primary dwarf galaxy. In particular, we adopted an NFW profile (see Eq. \ref{NFW}) for the primary dwarf in each pair and calculated the escape velocity (see Eq. \ref{escp}) at the radius of the extent of the HI profile (hence this is limited to the sensitivity of the data). We subsequently compared the escape velocity from each primary at this radius to the average velocity maps of each galaxy after subtracting the systemic velocities of the galaxy of interest (see values in Table \ref{tab:bound})\footnote{We here assume that the measured radial velocity of the gas is representative of the actual velocity the gas.}. 

\begin{figure}
\centerline{\includegraphics[width=\columnwidth]{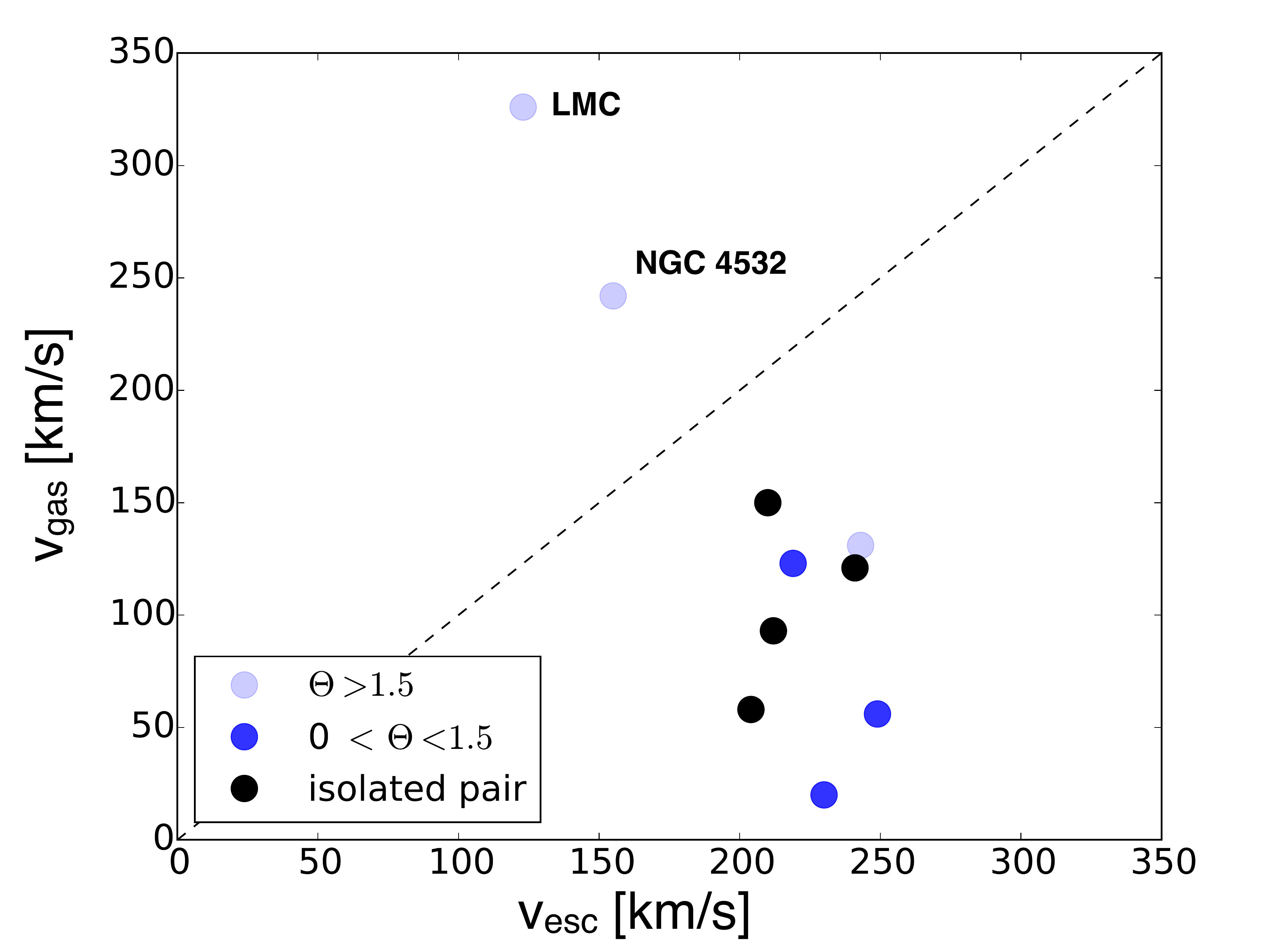}}
\caption{The gas velocity at the edges of the HI envelopes/profiles (estimated from velocity channel maps after subtracting the systemic velocity of the galaxy of interest) vs the escape velocity (see Eq. \ref{escp}) of the gas at this distance, calculated by adopting an NFW profile (see Eq. \ref{NFW}) for the primary dwarf in each pair. For all pairs except for the LMC pair and NGC 4532 pair, the extended gas remains bound to the dwarfs. }
\label{fig:vel}
\end{figure}

For all pairs, except for the LMC and NGC 4532 pairs, we found that the escape velocity from the primary was larger than the gas velocity at the edge of the HI distribution (see Figure \ref{fig:vel})\footnote{The mass of the secondary dwarf was not included in this calculation, but including it would only enhance the escape velocity further, which would lead to the same conclusion: that the gas is bound to the pairs.}. The LMC and NGC 4532 pairs are the only two systems with extended tails in the vicinity of a massive host. The escape velocities at the edges of each HI profile for all other pairs ranged from 204-249 km s$^{-1}$, where the velocity of the gas after subtracting the systemic velocities ranged from: 20-150 km s$^{-1}$ (see Figure \ref{fig:vel}). Most pairs have gas velocities which are 100 km s$^{-1}$ lower than the escape velocity. 
Hence, from this simple estimate the gas appears to remain bound to the dwarfs despite their tidal interactions, indicating that the gas in the outskirts will eventually fall back to the pairs and continue to fuel star formation. 

%%---------------------------------------------------------------------Figure----------------------------------------------------------------
%%---------------------------------------------------------------------Figure----------------------------------------------------------------

For the LMC/SMC pair, which has a long trailing stream, we estimated the escape velocity in the stream at a distance of 150 kpc from the LMC, though it has been found to extend to larger distances (e.g., \citealt{nide10}).
The velocity of the gas in the stream at this distance is 326 km s$^{-1}$ after subtracting the systemic (GSR) velocity of the LMC (\citealt{nide10}). Similarly, the escape velocity from the LMC at this distance is $v_\text{esc}$ = 123 km s$^{-1}$. The velocity limit of 123 km/s encompasses most of the HI mass (see \citealt{putman12}, Figure 2), but the extended ionized component of the stream extends well beyond this and is most likely unbound. 
For comparison, the escape velocity from the MW (assuming a total mass of 2 $\times 10^{12}$ and adopting an NFW profile) at a distance of 100 kpc is $v_\text{esc} \sim$ 432 km s$^{-1}$, hence the strength of the MW potential is much stronger than the strength of the LMC potential at the edges of the trailing stream. 

Similarly, the NGC 4532/DDO137 pair has a long trailing tail stretching 500 kpc away from the pair (\citealt{koop08}). The majority of the mass of the system is within 150 kpc, so in consistency with the LMC calculation above we estimate the escape velocity at 150 kpc from NGC 4532. We found that $v_\text{esc}$ = 155 km s$^{-1}$. The velocity of the gas in the stream at this distance after subtracting the systemic velocity of NGC 4532 is 242 km s$^{-1}$. Hence, the material at this distance and out to the edge of the stream (at $\sim$ 500 kpc) is not bound to NGC 4532. To compare this to the tidal field from the host, NGC 4570, we computed the escape velocity from NGC 4570 (where we estimate the mass from abundance matching, and adopt an NFW profile) at a distance of 392 kpc from the host (as the pair is already 242 kpc away and the stream material is 150 kpc further). We found that $v_\text{esc}\sim $ 340 km s$^{-1}$. Hence, the strength of the host potential is much stronger than the strength of the NGC 4532 potential in the trailing stream, as the case for the LMC pair.

Although, this is a simple way of estimating the relative tidal influences on the trailing streams, we can conclude that the trailing streams for the LMC and NGC 4532 are likely unbound from the dwarfs and are most likely bound to their massive host galaxies. This provides an interesting insight into how gas is truly removed from these systems, as the gas in all isolated and intermediate tidal index pairs appears to remain bound in the mutual interaction between the dwarfs in the pairs. Dwarf-dwarf interactions seem to be an efficient way to ``park'' HI gas at large distances, providing a continuous source of fuel for star formation.  However, it requires external environmental forces to ultimately cut off this gas supply channel and quench low mass galaxies (supporting the findings of \citealt{stierwalt15}), and the tidal interactions alone will not transform gas rich dwarf irregulars into gas poor systems, nor feed the IGM with significant baryons.   In the future, it will be interesting to take deeper HI observations of IC 2058 and other dwarf pairs as a further test.

\subsection{Exploring the Effects of the CGM of the Host Galaxies}\label{sec:cgm}
Since we think ram pressure could be important for the gas removal process (see Section \ref{sec:dens} and \ref{sec:bound}), we explore the influence from ram pressure further in this section. In particular, we make order of magnitude estimates of the required CGM densities of the host galaxies for ram pressure to explain the asymmetries in the HI profiles other than the bridges.

The COS-Halos Survey (\citealt{tum13}, \citealt{werk14}) found that more massive galaxies (logM$_*$/M$_{\odot}$$<$ 11.5) have halo gas within 160 kpc, which was the limiting distance from their central galaxies. Similarly, \citet{bordoloi14} found that a substantial amount of carbon is located $>$ 100 kpc away from galaxies of masses 9.5 $<$ logM$_*$/M$_{\odot}$$<$ 10, and \citet{liang14} found CIV enriched halos out to $\sim$ 160 kpc for galaxies of similar masses. Whether or not this CGM extends further from the centres of these galaxies is yet to be determined, however, theoretical work predicts that the CGM extends out to virial radius of galaxies (e.g. \citealt{joung12}, \citealt{hummels13}, \citealt{fuma14}).  

\begin{figure*}
\centerline{\includegraphics[width=\textwidth]{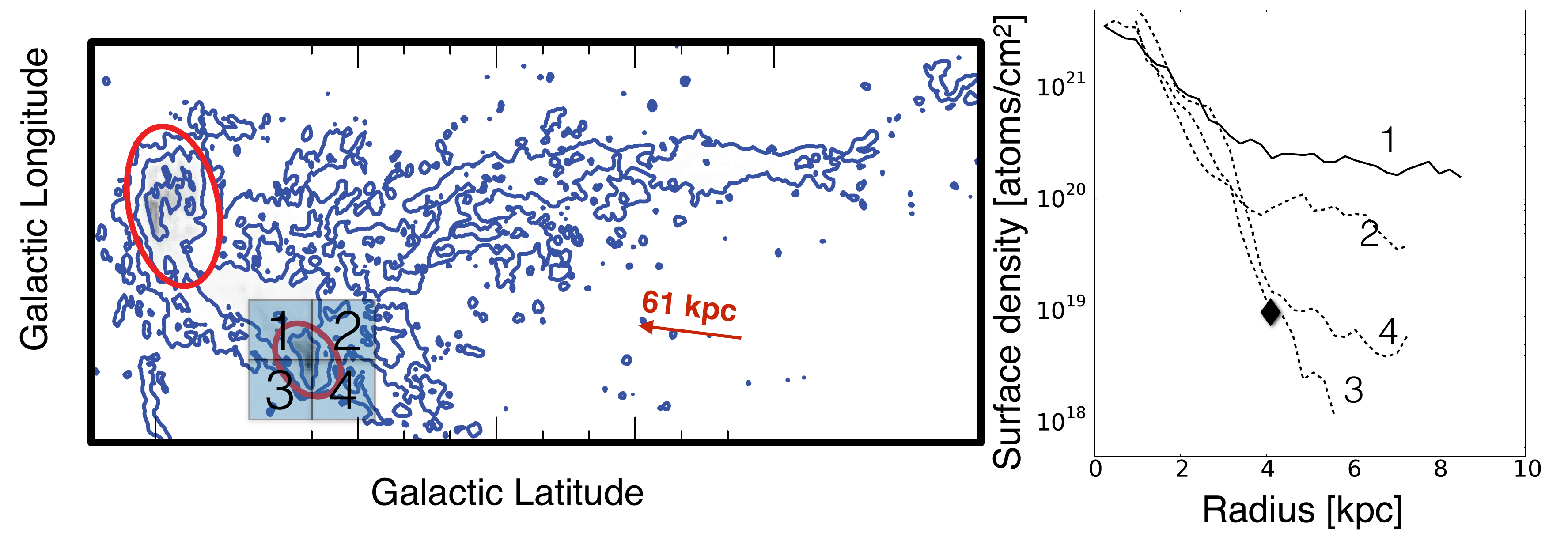}}
\caption{\textbf{Left}: HI map of the Magellanic System (see \citet{putman03} for Galactic coordinates of the data) with four regions (see numerated boxes) centred on the 2MASS ellipse of the SMC. The outer column density shown in the map is N(HI) = 1.0 $\times$ 10$^{19}$ atoms cm$^{-2}$.  
\textbf{Right:} Surface density profiles of the SMC HI distribution in four different directions (see numerated regions on map). The black diamond shows the radial extent at which the SMC disk is truncated ($R_{\text{trunc}}$ = 4.1 kpc). The surface densities vary by several orders of magnitude in column density in the four directions due to: the presence of a dense bridge connecting the LMC and SMC (solid line, region 1), the Magellanic Stream (2), the material lagging behind the SMC disk (4) and the general direction of motion towards the MW, where the profile is truncated (3). The sensitivity of the data is limited to a column density of N(HI) = 2.0 $\times$ 10$^{18}$ atoms cm$^{-2}$. }
\label{fig:smc}
\end{figure*}

One way to quantify the importance of ram pressure stripping is to use the \citet{gunn72} relation. \citet{salem15} demonstrated that they could probe the density of the halo material of the Milky Way using the \citet{gunn72} relation and by assuming that the one-sided truncation of the LMC HI disk was due to ram pressure by the Milky Way's halo. 
They found that the Milky Way halo density is 1.1 $\pm$ 0.44 $\times$ 10$^{-4}$ cm$^{-3}$ at the pericenter passage of the LMC (R = 48.2 $\pm$ 5 kpc). Interestingly, the high sensitivity data of the SMC shows large variations in column densities in the four different regions (see Figure \ref{fig:smc}), and we can examine the ram pressure effect on the SMC from this. In region 3 the surface density profile deviates from the other directions at a column density of $\sim10^{19}$ atoms cm$^{-2}$ at a radius of R = 4.2 kpc (see Figure \ref{fig:smc}). If we define R = 4.2 kpc as the truncation radius of the SMC based on the deviation in column densities at this radius, we can do an estimate of the halo density needed to produce this truncation radius using Eq. \ref{eq:voll}. Using the 3D velocity of the SMC (\citealt{kalli13}), we obtain a halo density of the MW of $\sim$ 6 $\times 10^{-5}$ cm$^{-3}$ (see Table \ref{tab:cgm}) at the distance of the SMC (61 kpc). This is somewhat lower than the \citet{salem15}, but reasonable for the Milky Way halo, especially given the potential partial shielding by the LMC. 

To estimate the required host halo densities to produce truncation features in all  $\Theta > 0$ primaries, we use the same procedure as for the SMC outlined above. We list the values used for our simplified Gunn \& Gott relation in Table \ref{tab:cgm}. For the HI surface density profiles that do not show rapid drops in column densities in the inferred direction of motion, we use the extent of the data as the truncation radius.

We first used the relative velocities of the pair to the hosts to get an upper limit on the CGM densities (as we here ignore any tangential motion and therefore underestimate the velocities). Note that we use the 3D velocities of the LMC and SMC (\citealt{kalli13}), which are higher than the line of sight $\Delta v$ for these galaxies ($\Delta v_{LMC} = 278$ km s$^{-1}$ and $\Delta v_{SMC} = 158$ km s$^{-1}$). Using line of sight $\Delta v$ for the LMC and SMC instead of the 3D velocities would yield slightly higher CGM densities, as the required CGM density scales inversely with velocity. 

Secondly, to estimate a lower limit on the CGM densities we assumed that the largest velocity they could move through the halos of the host galaxies would be the escape velocity from their host (although they could in principle be moving faster). For all $\Theta > 0$ pairs except for NGC 4490 this velocity was higher than the relative velocity of the pair and hosts. The values used and the results of the calculations of upper and lower limit on the CGM densities required for the halos are listed in Table \ref{tab:cgm}. 

Using the relative velocity of the host and pair for all  $\Theta > 0$ primary dwarfs, we find that the $\rho_{\text{CGM}}$ of their hosts need to be 3 $\times$ $10^{-4}$ - 2 $\times$ $10^{-5}$ cm$^{-3}$ at their current locations to explain the deviations in their profiles from a truncation due to ram-pressure.  Using the escape velocity from the host, we find that the $\rho_{\text{CGM}}$ of the hosts need to be 3.5 $\times$ $10^{-5}$ - 5.3 $\times$ $10^{-6}$ cm$^{-3}$ (here we leave out NGC 4490 as its $v_\text{esc} < v_\text{sep}$, see Table \ref{tab:cgm} for CGM densities required for this system). While the halo density result is reasonable for the LMC \& SMC\footnote{Note that we get lower values for $\rho_{\text{CGM}}$ than \citet{salem15}, due to the simplified version of the Gunn \& Gott criterion used here.}, it seems unlikely that the other pairs (see separate discussion of IC 2058 below) are moving through densities of  $10^{-4}$ - $10^{-5}$ cm$^{-3}$, since these dwarfs are $>$ 200 kpc from their hosts in projection. From theoretical predictions, the halo densities of MW type galaxies are $<10^{-5}$ cm$^{-3}$ at radii $>$ 200 kpc (e.g. \citealt{sommer06}), which is also supported observationally through studies of the Milky Way's hot halo gas (e.g. \citealt{miller13}) and by the COS-Halos Survey results finding declining metal surface density profiles with radius within 160 kpc of $L^*$ galaxies (\citealt{werk13}, \citealt{werk14}). Hence, ram pressure does not appear to be sufficient to truncate and remove gas from these systems (unless they are in fact moving at much larger velocities than assumed here). 
 
An exception is IC 2058. If it is moving at its escape velocity through the halo material of its host, the required halo density is $\sim 5 \times 10^{-6}$ cm$^{-3}$, which might be plausible at the projected distance of 91 kpc from its quite massive host, NGC 1553. The high inclination of the system makes it difficult to assess whether the HI profile is indeed truncated.  With higher resolution data we will be able to examine this and with deeper data we can see if trailing material is present in the opposite direction of its host.

For the NGC 4532 pair, it is possible that even though ram pressure is not the main driver of gas loss, it is shaping the gas distribution. Ram pressure is also not thought to be the main gas loss mechanism for the Magellanic System (\citealt{besla12}) and may be less significant than expected even for low mass dwarf systems (\citealt{emerick16}). Ram pressure by the CGM of the host may be shaping the NGC 4532/DDO137 tail after the dwarf-dwarf interaction ``loosened up'' (pre-processed) the gas. The densities this system is moving through could also be somewhat higher than expected at this radius due to the environment at the outskirts of the Virgo Cluster.

Another possibility is that tides from the hosts are removing gas from the dwarfs and truncating the HI profiles at their tidal radii. To investigate this, we estimate the tidal radius, $r_t$, of our high tidal index dwarfs using equation 8.91 from \citet{binney08}:
\begin{eqnarray}\label{rt}
r_t = \left(\frac{m}{3M}\right)^{1/3}r_p
\end{eqnarray}
where $m$ is the dark halo mass of the primary dwarf (see Table \ref{tab:bound}) , the pericenter distance, $r_p$, is estimated as the current projected distance from the dwarf to the host (see Table \ref{table:hosts}) and $M$ is the dark halo mass and stellar mass of the host enclosed within $r_p$ calculated using an NFW profile. We find that for all high tidal index dwarfs $r_t > 5.5 \times R_\text{trunc}$ (see Table \ref{tab:cgm}). Hence, tides from the hosts likely do not have a substantial effect on removing gas from the dwarfs. The fact that we do not see a leading stream for NGC 4532 further disfavors the idea that its trailing tail is formed due to tides from its host. However, once the gas has left the dwarfs through other mechanisms, tides from the host can be stronger than the tides from the dwarfs themselves and gas might be lost to the hosts (as we found in Section \ref{sec:bound}). 

Hence for all pairs, the process of moving gas to large radii, the formation of bridges and asymmetries are most likely due to a mutual interaction between the two dwarfs in the pair. Additionally, ram pressure from the MW appears to be truncating the HI profiles of the LMC and SMC, and for both the LMC and NGC 4532 pair, ram pressure from the massive hosts appear to be shaping their tails. Therefore, our results indicate that the CGM of this system likely extends out to distances $>$ 200 kpc, as models predict (e.g. \citealt{joung12}, \citealt{hummels13}, \citealt{fuma14}). This will be observationally probed with the future CGM$^2$ project (Werk et al., in prep.) as they plan to map the CGM of $L^*$ galaxies out to 3 $\times$ $R_{\text{vir}}$.

%%---------------------------------------------------------------------Figure----------------------------------------------------------------
%%---------------------------------------------------------------------Figure----------------------------------------------------------------

%-------------------------------------------------------Dense bridges-------------------------------------------------------
\subsection{The Dense Bridges}\label{sec:bridges}
Based on our definition of bridges as being continuous in HI column density and having a velocity gradient that smoothly connects one galaxy to the next, seven of our ten dwarf pairs have ``true'' bridges connecting the galaxies (see Figure \ref{fig:sample} and Table \ref{table:HI}). For the ones that do not have bridges, pair separations are large ($>$ 40 kpc). Thus HI bridges appear ubiquitous in dwarf pairs, regardless of environment and thus serve as the clearest hallmark of interaction. The column density in these bridges all have N(HI) $\gtrsim$ 10$^{20}$ atoms cm$^{-2}$, except for the bridge connecting NGC 4532 and DDO137, which has N(HI) $\sim 8 \times 10^{19}$ atoms cm$^{-2}$ (and is probed by a larger beam). These column densities are often several orders of magnitude higher than the column density at the same radial distance from the centre of the primary in the directions that are not pointing towards the companion dwarf (see solid lines in Figures \ref{fig:IsoEnv}, \ref{fig:InterEnv} and \ref{fig:DenseEnv}). 

Since our sample was selected specifically with the purpose of including interacting dwarf galaxy systems, it is not surprising that many of our pairs have bridges connecting the two dwarfs. However, the high HI column densities in the bridges not only hints at a recent close encounter between the pairs (e.g. \citealt{toomre72}, \citealt{combes78}, \citealt{hibbard95}, \citealt{barnes98}, \citealt{gao03}, \citealt{besla10}, \citealt{besla12})), but it also introduces a compelling case for why star formation should occur in the bridges connecting interacting galaxies and not in the trailing tails. \citet{besla12} showed that a high density bridge between the LMC and SMC can easily be reproduced from simulations of their mutual interaction. Interestingly, there is evidence for star formation in the Magellanic Bridge (e.g. \citealt{irwin85}, \citealt{demers98}, \citealt{harris07}) while no star formation has been observed in the Magellanic Stream.

From pure tidal theory, it is not expected that the density in the bridges should be higher than in the tails. However, the higher column densities in the bridges than in the tails can be explained from hydrodynamics (see e.g. hydro simulations in \citealt{besla10} and observational evidence of dense HI bridges in \citealt{gao03}), since ram-pressure acts most efficiently on low column density gas (i.e. produces low column density tails). 
In addition, the high density in the bridges could be explained from overlapping gas from both galaxies which could be occurring in a close encounter (this is likely the formation mechanism of the Magellanic bridge; \citealt{besla12}).

Several other studies have found evidence for star formation in HI bridges connecting more massive interacting galaxies (e.g. \citealt{mello08}, \citealt{condon02}), and evidence for star formation in bridges of pre-merger dwarf galaxies (including the NGC 4490/85 bridge) have been found in the UV (\citealt{smith10}). Studying the bridges of dwarf pairs in the optical, UV and higher resolution HI could reveal higher column density cores that have formed in gaseous tidal features, which could provide insight to a different mechanism for star formation than in disks (e.g. \citealt{werk10}).

\subsection{Star Formation in the Dwarfs}\label{sec:sfr}
Gas outflows from galaxies owing to supernova feedback is a different gas removal process that could move gas out to large distances (e.g. \citealt{hopkins14}, \citealt{arraki14}, \citealt{pontzen12}, \citealt{tolstoy09}, \citealt{clemens02}). Furthermore, it has been proposed that mergers between gas rich dwarfs can trigger starbursts and potentially transform the dwarfs into gas poor systems (e.g. \citealt{bekki08}). Hence, it is important to assess the impact of star formation as we are investigating extended gas around dwarfs and their mutual interactions. 

\citet{stierwalt15} showed that the dwarf galaxy pairs in their sample had enhanced star formation rates when at smaller pair separations. Furthermore, they showed that the dwarf galaxies that were star bursting (H$\alpha$ equivalent width: EQW $>$ 100 \AA) did not appear to be gas depleted. This agrees with the findings of \citet{bradford15} that the smallest gas fractions of dwarfs in isolation are $f_{\text{gas}} \sim$ 0.3, indicating that star formation and stellar feedback from e.g. supernovae does not remove all gas from these systems. Interestingly, the dwarf galaxies in the \citet{stierwalt15} sample that were gas depleted (hence within 200 kpc of a massive host) did not show an enhancement in SF with interaction stage (Stierwalt et al., in prep.). This indicates that once the gas is ultimately removed by the environment, interactions between dwarf galaxies do not enhance the star formation. 

\begin{figure}
\centerline{\includegraphics[width=\columnwidth]{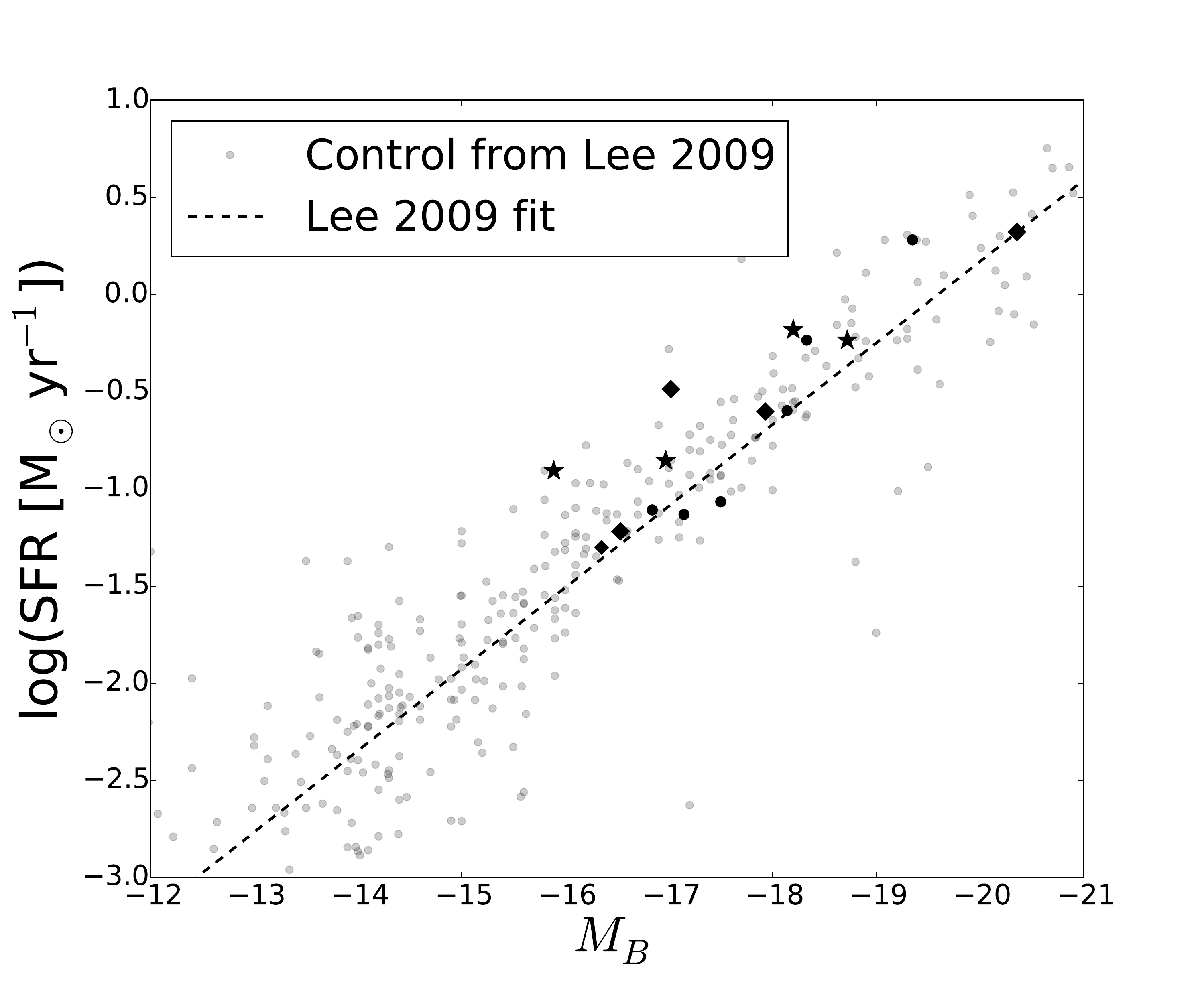}}
\caption{H$\alpha$ inferred star formation rates vs B-band magnitude for 300 dwarf galaxies from the \citet{lee09} sample (grey) plotted along with our sample of dwarf galaxies (black stars: EW $>$ 70 \AA, black diamonds: no EW available, black circles: EW $<$ 70 \AA). The dotted line represents the fit to the 300 dwarfs in the \citet{lee09} sample.}
\label{fig:lee}
\end{figure}

Motivated by gas outflows owing to supernova feedback and by the results in \citet{stierwalt15}, we investigate whether our paired dwarfs are outliers in terms of their SFs when compared to a larger sample of non-paired dwarfs. In particular, we compare the SFRs derived from H$\alpha$ luminosities of our dwarfs (see Table \ref{table:prop}) to the SFRs in \citet{lee09} (hereafter L09). The L09 sample consists of 300 dwarf galaxies (both pairs and single dwarfs) within 11 Mpc of the MW in a B-band luminosity range of -11 $<$ $M_B$ $<$ -20. Starburst galaxies are 6\% of this sample. Our SFRs were inferred from the H$\alpha$ luminosities based on the following relation:

\begin{eqnarray}\label{eq:sfr}
\text{SFR[M}_{\odot} \text{yr}^{-1}] = 7.9 \times 10^{-42}L(H\alpha)\text{(erg/s})
\end{eqnarray}
In Figure \ref{fig:lee} we show the SFRs of our dwarf galaxies (black points) over plotted on the L09 sample (grey points). Nine of our individual dwarf galaxies are overlapping with the L09 sample. L09 used a different relation to calculate the SFRs than Equation \ref{eq:sfr}, however we re-calculated all SFRs in their sample using Equation \ref{eq:sfr}. It is evident that our dwarfs are all within the scatter of the L09 sample, although a fit through our sample would yield a systematically higher SFR for a given $M_B$, than the L09 fit (dotted line). It is important to note that in the specific $B$-band range of our sample, the majority of the L09 data points also fall above their fit, which is based on a range of B-band absolute magnitudes from -11 $<$ $M_B$ $<$ -20. Seeing an enhancement in the star formation rates will depend on the interaction history of the pairs, hence they could have had a burst of star formation previously. The fact that our pairs do not appear to have an overall enhancement in SF when compared to the other L09 data in that $M_B$ band range, supports the idea that the extended gas is due to tidal pre-processing of the gas and not due to outflows from star formation.

To investigate whether it is reasonable that our dwarf pairs do not appear to have an overall enhanced SF, we can compare the results to similar studies of more massive galaxies. Recently, \citet{knapen15} found that while the majority of interacting (massive) galaxies do not have enhanced SFRs, those that do have extremely high SFRs when compared to their non-paired counterparts (see also \citealt{knapen2015}). \citet{patton13} found that the strongest enhancements for massive galaxies were seen at small separations: $<$ 20 kpc, which is in agreement with the results of \citet{scudder12} and \citet{davies15}, who also find that galaxy interactions trigger enhanced star formation. 

Whether dwarf galaxies are also more extremely elevated in their SFRs at separations $< 20$ kpc is unclear, as \citet{stierwalt15} did not look into pair separation bins smaller than 50 kpc. 
There are four dwarf pairs in our sample that have pair separations $<$ 20 kpc for which we have SFRs (the LMC pair, the NGC 4618 pair, the NGC 4490 pair and the NGC 672 pair, Table \ref{tab:local}). While their H$\alpha$ derived SFRs are within the scatter of the L09 sample (see Figure \ref{fig:lee}),
the H$\alpha$ equivalent widths (EQWs) of NGC 4490/85 (our closest pair) are both $>$ 65 \AA~ (\citealt{kenni08}), which could indicate that they have triggered star bursts (indeed L09 found that NGC 4449 and NGC 4485 are starbursts as their EQWs exceeded the logarithmic mean of their dwarf sample by 2$\sigma$). 
However, there are no signs that the four pairs with the smallest pair separations in our sample have systematically elevated SFRs compared to the other dwarf pairs. Hence, so far our findings do not agree with the results for more massive galaxies, but given the small sample size the SFRs of our dwarfs could be consistent with the type of enhancement seen for massive galaxies (e.g. \citealt{patton13}).

%---------------------------------------------------------------------Figure----------------------------------------------------------------
%---------------------------------------------------------------------Figure----------------------------------------------------------------

%---------------------------------------------------Conclusion-----------------------------------------------------------------
\section{Conclusion} \label{sec:conclusion}
In this paper we have investigated HI synthesis maps of a local sample of interacting dwarf galaxy pairs (the LV-TNT sample) and have considered the importance of environment and dwarf-dwarf interactions in removing and shaping the gas. We note that our sample consists of only 10 pairs and that our conclusions can be strengthened with future HI observations of interacting dwarf pairs. 

Our results and conclusions are summarized as follows:
\begin{itemize}
\item[1.]{Dwarf galaxy pairs residing in the proximity of a massive host galaxy ($\Theta > 1.5$) are affected morphologically by their environment, when compared to dwarf galaxy pairs interacting in isolation. Their surface density profiles are highly asymmetric indicating both tidal interactions with each other (bridges) and interactions with the halos of the massive hosts (extended tails in the directions of their host). In contrast, dwarf galaxy pairs with $\Theta < 1.5$ primarily have variations in their surface density profiles in the direction towards their companion, where dense bridges are present.}\\

\item[2.]{The majority of dwarf galaxy pairs have a large amount of gas in their outskirts ($>50\%$ of their total gas mass is beyond their 2MASS stellar extents) and they appear to have extended dense gas envelopes when compared to non-paired dwarf irregulars. 
This implies that dwarf-dwarf interactions move gas to the outskirts of these galaxies.}\\

\item[3.]{The gas remains bound to the systems from tidal interactions between the dwarfs alone. As such, this gas will be reaccreted by the system, providing fuel for future star formation. }\\

\item[4.]{Two of the three $\Theta >1.5$ pairs (in the vicinity of a massive host galaxy) have long, trailing tails. We find that the gas in these tails is unbound and likely lost from the pairs to the massive host halos.  This combined with point 3.~above indicates that the environment is what ultimately quenches low mass galaxies, by preventing the gas from returning to these systems.  This supports the conclusions from \citet{stierwalt15} and \citet{bradford15}.  }\\

\item[5.]{For the pairs in the vicinity of a massive host, ram pressure stripping alone is found to be insufficient to remove gas from the pairs.  Pre-processing via dwarf-dwarf interactions is key to enabling this gas supply channel to the CGM if they eventually fall into the halos of more massive galaxies. For the LMC and NGC 4532 pairs, ram pressure stripping appears to be shaping the gas once it leaves the pairs.}\\

\item[6.]{Seven of our ten dwarf pairs have dense bridges connecting them. The bridge column densities are higher than in other regions in the outskirts of the systems, potentially explaining why SF can be ongoing in the bridge and not in the other extended structures (e.g. as observed in the LMC/SMC and NGC4490/85 bridges). Such high column density tidal structures require close passages between the two pairs to form.}\\

\item[7.]{The star formation rates of our sample are within the scatter of the \citet{lee09} sample of nearby dwarf irregular galaxies, hence our dwarf pairs do not have substantially elevated star formation rates. This supports that the extended structures we find are from tidal interactions and not from outflows.}\\
\end{itemize}

This study highlights dwarf-dwarf interactions as an important part of the baryon cycle of low mass galaxies, enabling the ``parking'' of gas at large distances to serve as a continual gas supply channel until accretion by a more massive host prevents this gas from being reaccreted by the pair.

\section*{Acknowledgements}
We thank Fiona Audcent-Ross, Marcel Clemens, Thijs van der Hulst, Diedre Hunter, Rebecca Koopmann, Janice Lee, Moire Prescott, Eric Wilcots and Dennis Zaritsky for sharing their data of the dwarfs in this sample. We also thank Jessica Werk, Gerhardt Meurer, Lauranne Lanz, Joshua Peek and Kathryn V. Johnston for useful discussions. MEP acknowledges support for program 13383 from NASA through a grant from the Space Telescope Science Institute, which is operated by the Association of Universities for Research in Astronomy, Inc., under NASA contract NAS5-26555. XF is supported by an NSF Astronomy and Astrophysics Postdoctoral Fellowship under award AST-1501342. We have made use of the WSRT on the Web Archive. The Westerbork Synthesis Radio Telescope is operated by the Netherlands Institute for Radio Astronomy ASTRON, with support of NWO. This paper includes archived data obtained through the Australia Telescope Online Archive (http://atoa.atnf.csiro.au). The Australia Telescope Compact Array (/ Parkes radio telescope / Mopra radio telescope / Long Baseline Array) is part of the Australia Telescope National Facility which is funded by the Australian Government for operation as a National Facility managed by CSIRO.

\appendix
\section{Dwarf Galaxy Pairs in our Sample}\label{sec:samp}
In this appendix, we describe each dwarf galaxy pair in our sample.  The pairs are listed in order of decreasing tidal index, $\Theta$. See Table \ref{tab:local} and \ref{table:prop} for details on the physical properties of each pair, and Table \ref{table:HI} and Figure \ref{fig:sample} for details on all HI observations. 

Using Eq. \ref{NFW}, Eq. \ref{escp} and the line of sight velocity separation of each pair listed in Table \ref{tab:local}, we find that each secondary dwarf is bound to its primary dwarf (i.e. $v_\text{sep}$ $<$ $v_\text{escape}$).

\subsection{LMC, SMC}
As described in Section \ref{sec:intro}, the evidence for an ongoing interaction between the LMC and SMC is the extended HI distribution surrounding the pair (e.g. \citealt{putman03} which is the data set used in this study). The Magellanic Clouds are currently $\sim$ 23 kpc apart (11 kpc projected) and 50 kpc from the Milky Way disk. Their currently measured stellar mass ratio is (M$_\text{*LMC}$/M$_\text{*SMC}$) $\sim$ 10 (\citealt{marel02}, \citealt{stani04}). As of yet, stellar components to the gaseous streams have not been found, and therefore the interaction is revealed only by the gaseous extensions (although \citet{bel15} recently found an over density of Blue Horizontal Branch stars approximately aligned with the proper motions of the clouds). The Magellanic System has an HI mass $>$ 10$^9$ M$_{\odot}$ and an extended ionized gas mass of the same magnitude (\citealt{putman03}; \citealt{fox14}). 

The origin of the extended gaseous features has remained ambiguous since their discovery (\citealt{mathewson74}). Many models have invoked primarily tidal or ram pressure forces from the Milky Way halo to create the gaseous features (\citealt{gardiner96}, \citealt{connors06}, \citealt{mastropietro05}, \citealt{diaz11}); others create the gas streams primarily through the interaction of the Clouds themselves (\citealt{besla10}, \citealt{besla12}), while including the tidal influence of the MW.  The Clouds possess a well studied bridge that has a smooth velocity gradient of 50 km s$^{-1}$ (\citealt{nide10}), which is evidence for a mutual interaction between the two dwarfs. The bridge is known to be an ongoing site of star formation (SF) (e.g. \citealt{irwin85}, \citealt{demers98}, \citealt{harris07}).

The Magellanic Clouds are important members of our Local Volume sample, as they are in a high tidal index environment ($\Theta = 3.7$) owing to their proximity to the MW.

\subsection{IC 2058, PGC 75125}
IC 2058 and its companion PGC 75125 are located 18.1 Mpc (\citealt{nasonova11}) from the Milky Way, but only 91 kpc in projection from the more massive galaxy: NGC 1553. The pair's close proximity to a more massive galaxy, results in a high tidal index of $\Theta = 3.2$. Observations of this system were done in 2006 and 2007 with the Australian Compact Telescope Array (ATCA) and are presented for the first time in this work (see Figure \ref{fig:sample} and Table \ref{table:HI})\footnote{The data were uniformly weighed with a spectral resolution of 20 km s$^{-1}$ and pixel size of 19$\arcsec$/pixel.}. \citet{kilborn05} studied the system using single dish observations from the Parkes Telescope and concluded that it is a part of the NGC 1566 group (the projected distance to NGC 1566 is 350 kpc). The two dwarf galaxies, IC 2058 and PGC 75125, are separated from each other by only 9.5 kpc in projection, and a small HI bridge is connecting them (see Figure \ref{fig:sample}). The bridge connecting the two galaxies is a strong indicator of an ongoing tidal interaction (e.g. \citealt{toomre72}). Additionally, the line of sight velocity separation of the two dwarfs is only $\sim$ 10 km s$^{-1}$, and their stellar mass ratio is quite high ($\sim$ 11, similar to the Magellanic Clouds). Hence, it is likely that the two galaxies are in fact interacting.

H$\alpha$ imaging of the system (available on NED) suggests a potential ram pressure stripping of the smaller companion, as both edges of the disk seem to be warped in the same direction away from the larger galaxy. However, due to the small angular size of PGC 75125, this can not be confirmed with the HI map. 

\subsection{NGC 4532, DDO 137}\label{sec:N4532}
NGC 4532 and DDO 137 are defined as Magellanic class dwarf galaxies and are interacting, as evident from their large common HI envelope and HI tail that extends for 200-500 kpc (\citealt{koop08}; \citealt{hoffman92}). These galaxies have elevated star formation rates and a disturbed central kinematic structure (\citealt{koop04}; \citealt{hoffman99}). The two galaxies are separated in projection by 48 kpc, their velocity separation is only $v_{\text{sep}}\sim 27$ km s$^{-1}$, and they have a similar stellar mass: (M$_\text{*NGC 4532}$/M$_\text{*DDO137}$) = 1.4. The system is located in the outskirts of the Virgo Cluster 13.8 Mpc from the MW (\citealt{tully09}), and it is located $\sim$ 240 kpc away from NGC 4570 in projection, which has M$_{*} > 4 \times 10^{10}$ M$_{\odot}$. This results in a tidal index of $\Theta = 1.5$, which places the pair in our high tidal index group (see Table \ref{table:hosts}). However, if the NGC 4532 pair is farther from the MW than 13.8 Mpc (as suggested by e.g. \citealt{willick97}), the pair would be in the intermediate tidal index group (see Table \ref{table:hosts}) and could be less affected by its host than assumed throughout the rest of the paper. 

The large trailing tail (see Figure \ref{fig:sample}) could be an indication of ram pressure stripping by the more massive NGC 4570. Due to the fortunate alignment of a bright background quasar with the NGC 4532/DDO137 tail, the Hubble Space Telescope Cosmic Origin Spectrograph (HST-COS) was used to probe the metallicity and ionization conditions of the tail (HST Proposal ID: 13383). However, little or no absorption was detected at the velocity of the system despite going through the HI tail. This little or non-detection could indicate that there is no ionized gas present in the trailing tail of the system. In contrast, we know that the Magellanic Stream has a large amount of ionized gas (see \citealt{fox14}). The fact that NGC 4532/DDO 137 is located much farther from its host than the Magellanic System is from the MW, might explain why its tail is not ionized (see Section \ref{sec:cgm} for more details). 

\subsection{NGC 4618, NGC 4625}
The NGC 4618 and NGC 4625 dwarf galaxy pair is an example of two Magellanic spirals that have adjoining HI distributions (see Figure \ref{fig:sample}). The pair is located 7.9 Mpc from the MW and are only separated by 9.2 kpc in projection and have a $v_{\text {sep}} = 77$ km s$^{-1}$. They reside in the vicinity of the more massive galaxy, Messier 94, which is $\sim$ 240 kpc away from the pair, yielding a tidal index of $\Theta = 1.4$ (i.e. much lower than the MW's influence on the Magellanic System (MS)). The morphology of the HI distribution around NGC 4618 reveals a ring-structure ($\sim$10\% of total HI mass), which could indicate recent tidal interactions. However, \citet{kac12} argued that it is not clear that this is due to close proximity of NGC 4625. In fact, they conclude that the two Magellanic spirals are not interacting. This was also argued by \citet{bush04}, who pointed out that the degree of asymmetry in both galaxies is indistinguishable from the expected asymmetry in lopsided galaxies. From the maps presented in \citet{hulst01} the HI column density appears continuous between the two galaxies. However, \citet{kac12} pointed out that the interface where the two disks overlap are separated by 50 km s$^{-1}$ in velocity (i.e. a non-smooth velocity gradient), and that it therefore unlikely that this is a true gas bridge (see their Figure 13).

However, due to the disturbed velocity field of NGC 4618 (see \citealt{bush04}) and the fact that the two galaxies are in close proximity of one another in both velocity and position space, with overlapping HI disks, we include them in our sample. We note that the kinematics and HI distribution of NGC 4625 are surprisingly regular, and we consider this system as a potential non-interacting dwarf pair in the rest of the paper.

\subsection{NGC 4490, NGC 4485}
The NGC 4490 and NGC 4485 pair is an example of a well studied, nearby Magellanic System analog. The two dwarfs are surrounded by one of the most extended known HI envelopes ($\sim$ 50 kpc), and they are connected by a dense HI bridge, which is continuous in its HI column density and has a smooth velocity gradient (\citealt{huch80}, \citealt{clemens98}, and see Figure \ref{fig:InterEnv}). The dwarfs are separated by only $\sim$ 7.5 kpc, and their $v_{\text {sep}} = 72$ km s$^{-1}$. Their interaction is also evident from the disturbed stellar morphology of NGC 4490 (warped disk), and from the fact that NGC 4485 is experiencing a starburst (\citealt{lee09}). While \citealt{clemens98} argue that the envelope is formed from outflows, \citet{elmegreen98} argued that both NGC 4490 and NGC 4485 have stellar tidal tails associated with their disks, which strengthens the argument for a tidal origin of the HI envelope. 

The closest massive galaxy to the dwarf system is NGC 4369 (M$_{*} = 2.6 \times 10^{10}$ M$_{\odot}$), which is located $\sim$ 310 kpc from the pair in projection (although $v_{\text {sep}}$ $>$ 400 km s$^{-1}$ for the pair and host). It has been discussed whether the extended envelope is of tidal origin or due to ram pressure. However, as NGC 4369 is $\sim$ 10 times less massive than the Milky Way and as the dwarf pair system is quite far from the more massive galaxy, a ram pressure stripping scenario seems unlikely (see Section \ref{sec:cgm}). The pair is therefore a nice analog to the Magellanic System prior to infall, as it has an extended, more symmetric HI distribution surrounding the pair. 

In this paper we adopt the distance of. 7.14 Mpc to the pair (\citealt{theureau07}), while e.g. \citet{kara13} found a distance of 5.8 Mpc to the pair. Adopting a different distance would yield slightly different stellar masses, physical pair separations and inferred physical extents of the galaxies which will not affect our conclusions of this paper (see how these parameters scale with distance in Table \ref{tab:local} and \ref{table:prop}). Using the distance found in \citet{kara13}, the estimated projected distance to the host would be 256 kpc. However this would not change the tidal index group that we use for this pair throughout the paper (see Table \ref{table:hosts}).

\subsection{ESO 435-IG16, ESO 435-IG20}
Observations of ESO 435-IG16 \& ESO 435-IG20 were taken with ATCA in 2002 (\citealt{kim15}). The two dwarfs are located 11.6 Mpc from the Milky Way and are at a large separation from each other ($\sim$ 100 kpc). Despite the separation, evidence of an interaction or ongoing gas removal between the two galaxies is seen in the eastern extension of ESO 435-IG16 and in the tail of ESO 435-IG20  (see Figure \ref{fig:sample}). Their stellar mass ratios are 1:10, and they are only separated by 9 km s$^{-1}$ in line of sight velocities. The pair is located $\sim$ 860 kpc from a more massive host (NGC 2997), which is $\sim$ 2 times farther than the $R_{200}$ of the host. The pair is also located $\sim$ 220 kpc in projection from another dwarf (NGC 3056) which is on the West (right) side of the pair. Interestingly, the HI distributions seem more extended to the opposite direction of the location of NGC 3056, however it is unlikely that dense halo material of this dwarf extends out to 220 kpc (\citealt{bordoloi14}), which is beyond R$_{200}$ of NGC 3056. HIPASS observations of the system (which reach a column density of N(HI) $\sim 10^{18}$ cm$^{-2}$), show that the two dwarfs are covered by a large envelope of neutral gas  surrounding both galaxies (\citealt{kim15}).

\subsection{NGC 3448, UGC6016}
The NGC3448 and UGC6016 pair is a part of the M81 group of galaxies and is located  24.7 Mpc from the MW. The closest galaxy more massive than 5 $\times 10^9$ M$_{\odot}$ is located within a projected distance of $\sim$ 1050 kpc from the pair (NGC 3310), yielding a tidal index of $\Theta < 0$. Hence, we classify this pair as being isolated. Both a leading and trailing stellar tail of NGC 3448 have been observed with multiple telescopes (first published in the ARP catalog 1966). The two dwarf galaxies are separated by $\sim$ 30 kpc and are rotating with opposite spin. Their line of sight velocities are separated by 143 km s$^{-1}$. Models of their interactions indicate that this is a retrograde encounter (hence the rotation of UGC6016 is opposite to that of the orbit of the encounter between the two galaxies). \citet{toomre72} showed that a retrograde encounter between galaxies will not lead to a dramatic distortion in the kinematics (see also \citealt{donghia09}). This explains the relatively undisturbed, regular stellar dynamics of UGC 6016 as discussed by \citet{noreau86}. \citet{noreau86} showed that the perturbed HI morphology of NGC 3448, could be satisfactorily reproduced through a tidal interaction scenario between the two  galaxies. \citet{bertola84} studied this system in detail using optical, UV and radio observations. They also conclude that a tidal interaction can explain the morphology of the gas distribution surrounding the pair, which is overlapping (the peak of the HI emission is centred on NGC 3448, and the other peak appears to be offset from the optical centre of UGC 6016) with a smooth velocity gradient between the pairs.

\subsection{UGC 9562, UGC 9560}
Compelling evidence that the dwarf galaxies UGC 9562 and UGC 9560 are interacting was presented by \citet{balkowski78} and \citet{cox01}, who showed that there is a gaseous bridge connecting the two galaxies (M$_{\text{HI}}$ = 2.8 $\times 10^{8}$ M$_{\odot}$). The bridge is continuous in HI column density between the two pairs, and the velocity gradient is smooth (see \citealt{cox01} for velocity map and higher resolution HI data). The system is located 25.5 Mpc from the Milky Way and the two pairs are at a projected separation of $\sim$ 33 kpc and a line of sight velocity separation of 112 km s$^{-1}$. In addition to the gaseous bridge connecting the two dwarfs, \citet{cox01} found evidence of a polar ring around UGC 9562 with H$\alpha$ line emission, which could be a remnant of a recent interaction. The two galaxies are located more than 1.5 Mpc from any massive galaxy, hence they are evolving in isolation. 

\subsection{NGC 0672, IC 1727}
NGC 672 and IC 1727 are two dwarf spiral galaxies, that are connected through a massive HI bridge (see Figure \ref{fig:sample}). The bridge is continuous in HI column density, and the velocity gradient in the bridge connecting the two galaxies is smooth (see WHISP velocity maps: \citealt{hulst01}). They are located 7.9 Mpc from the MW and their optical centres are separated only by 19 kpc. The galaxy pair is evolving in isolation, as the nearest, more massive galaxy is $>$ 1.5 Mpc away. \citet{combes80} argued that a gravitational interactions between the two galaxies is taking place, based on the offset in HI centres compared to the galaxies optical centres, and from the bridge of gas connecting them. Subsequently, \citet{ramirez14} found that the interstellar medium of IC 1727 is very perturbed, which could be a sign of recent interaction. Furthermore, the WHISP map  (\citealt{hulst01}) and the more recent HALOGAS map (\citealt{heald11}) of the system shows a tidal arm trailing behind NGC 672, which also suggests an ongoing tidal interaction.

\subsection{NGC 4449, DDO 125 (and halostream)}\label{sec:N4449}
NGC 4449 is unusual since it has two counter rotating gas systems (\citealt{hunter98}) and has a large tidal or spiral feature surrounding its optical disk. DDO 125 is located $\sim$ 40 kpc from the centre of NGC 4449 with a velocity separation of only 10 km s$^{-1}$. The two galaxies are both located approximately 4 Mpc from the Milky Way, and there is no bridge connecting the two galaxies, though an extension in HI is seen in the direction of DDO 125 from the NGC 4449 HI distribution. DDO 125 does not seem tidally disturbed in HI nor in its stellar component, however its $M_{\text{HI}}$/$L_B$ is low, indicating that it could have lost a substantial amount of gas in an encounter with NGC 4449. 

\citet{kara07} noticed an elongated stream candidate near NGC 4449, and recently \citet{delgado12} presented deep, wide-field optical imaging of the faint stellar stream (NGC 4449B) 10 kpc southeast of NGC 4449, which has a stellar mass of only $\sim$5 $\times$ 10$^7$ M$_{\odot}$. Additionally, \citet{rich12} pointed out the ``s''-shape of the stream, which is characteristic of an ongoing tidal disruption. This stellar stream provides an alternative explanation for the complex HI structure of NGC 4449 (which includes rings, shells and a counter-rotating core). The present day mass ratios of NGC 4449 to DDO125 and NGC 4449 to NGC 4449B are 15 and 80, respectively. There is no massive host galaxy in the vicinity of the pair, which is therefore evolving in isolation.

\begin{table*}
\caption{Local interacting dwarf galaxies}
\label{tab:local}
\begin{tabular}{ccccccc}
\hline
Primary &Secondary & Distance &Pair sep., $\Delta$d$^j$ &Velocity sep., $\Delta$v$^k$& Prim. stellar mass  & Sec. stellar mass\\[2pt]
name & name &[Mpc] & [kpc]$\times (D/D_{\text{Table1}})$&[km s$^{-1}$] & $\times 10^9$[M$_{\odot}$]$\times (D/D_{\text{Table1}})^2$& $\times 10^9$[M$_{\odot}$]$\times (D/D_{\text{Table1}})^2$ \\
\hline
LMC 	& 	SMC 			& 	0.05$^a$/0.061$^b$ & 11 & 120		&2.3	&0.23	\\[2pt]
IC 2058 	&	PGC 75125 		 &	18.1$^c$ 			& 		9.5	& 10 & 		2.7	&0.24 		\\[2pt]
NGC 4532&	DDO 137 			&	13.8$^d$ 			&  		48 & 27    & 		6.2	&3.0			 \\[2pt]
NGC 4618&	NGC 4625 		&	7.9$^e$		&			9.2 & 77    &		4.3	&1.3			  \\[2pt]
NGC 4490&	NGC 4485 		 &	7.14$^f$			&		7.5   & 72    		&7.2	&		0.82	 \\[2pt]
ESO435-IG16	& ESO435-IG20 	&	11.6$^g$			&		 101  &9.0   &		2.3	&0.28 		\\[2pt]
NGC 3448& UGC 6016		 	&	 24.7$^g$ 			& 			29  & 143   &		3.6$^l$ 	&0.081$^l$ 	 	\\[2pt]
UGC 9562&	UGC 9560 		&	 25.5$^g$ 				&			33  & 112   &  	2.0		&	1.0	 \\[2pt]
NGC 672&IC 1727				 &	7.9$^h$ 			&			  19 &84    &		5.0	&0.96	\\[2pt]
NGC 4449& DDO 125			 &	3.82$^i$ 				& 		40 &   10 &		3.7	& 0.24	 \\[2pt]
\hline
\multicolumn{7}{l}{a: \citet{pietrzy13}, b: \citet{cioni00}, c:, \citet{nasonova11}, d: \citet{tully09}}\\
\multicolumn{7}{l}{e: \citet{kara13}, f: \citet{theureau07}, g: NED kinematic flow distance (Virgo + GA + Shapley) }\\
\multicolumn{7}{l}{assuming $H_0$ = 73 km/s/Mpc, corrected by \citet{mould00}. h: \citet{sohn96}, i: \citet{anni08}}\\
\multicolumn{7}{l}{j: Based on angular separation from centre to centre and converted to physical distance based on distance to pair.}\\
\multicolumn{7}{l}{k: From NED redshift: $\Delta$v =  c $\times | z1 - z2 | /  ( 1 +  (z1 + z2)/2) .$}\\
\multicolumn{7}{l}{l: \citet{lanz13}}\\
\end{tabular}
\end{table*}

%
%%-----------------------------------Dwarf properties-----------------------------------
%
\begin{table*}
\caption{Properties of dwarf pairs in our sample}
\label{table:prop}
\begin{tabular}{@{\:}c@{\:}c@{\:}c@{\:}c@{\:}c@{\:}c@{\:}c@{\:}c@{\:}}
\hline
Host dist.& Dwarf Pair & r$_{\text{ext}}$$^{**}$ &r$_{\text{ext}}$$^{***}$& DHI$_{\text{ext}}$$^j$ &DHI$_{\text{ext}}$ & SFR prim.& SFR sec. \\
&  &[arcsec]  &[kpc] &[arcsec]  &[kpc]  & [M$_{\odot}$yr$^{-1}$]  & [M$_{\odot}$yr$^{-1}$] \\
  &  &  &$\times (D/D_{\text{Table1}})$&  &$\times (D/D_{\text{Table1}})$  & $\times (D/D_{\text{Table1}})^2$ & $\times (D/D_{\text{Table1}})^2$\\
\hline
0-500 &LMC/SMC 		 &27996/16634&	6.8/5.0	&42988.0/22721& 10/6.7	& 0.25$^a$&0.05$^b$\\
&IC 2058/PGC 75125	 &96.7/16&		8.5/1.4	&320/-	&	28/-& -  &-\\
&NGC4532/DDO137 	& 88/52*& 		5.9/3.5 		&580/550	&	39/	37&  0.71$^d$&-\\
&NGC4618/25 & 		153/74& 			5.3/2.8 		&480/500	&	18/19	& 0.29$^f$&0.037$^f$ \\
&NGC4490/85  &		 213/74 & 			7.4/2.6 		&500/710&	17/25	&2.6$^e$&0.17$^e$\\
&ESO435-IG16/IG20 	&62/19& 			3.5/1.1 		&325/180&	18/10	&-&0.012$^c$\\

\\1000 - 1200 &NGC3448/UGC6016 & 115/55* &  14/6.6	&750/396 &	90/47& 0.60$^g$&0.75$^j$ \\

\\$>$1500&UGC 9562/60 	&  23/17 &			 2.8/2.1			&220/114&27/14		& 0.085$^h$&0.40$^h$\\
&NGC672/IC1727		 &	 220/115&		 8.4/4.4 	&790/640&	30/25	 &  0.24$^i$&0.11$^i$\\
&NGC4449/DDO125 	&	240/133.5 & 	4.4/2.5 		&1114/-&	20/-	& 0.59$^e$&- \\
\hline
\multicolumn{8}{l}{$^*$ Indicates that 2MASS observations of the galaxy were not available, in which case the r-band extent was used.}\\
\multicolumn{8}{l}{$^{**}$ All r$_{\text{ext}}$ are determined from the 2MASS scale lengths (see Section \ref{sec:prop}).}\\
\multicolumn{8}{l}{$^{***}$ Same as r$_{\text{ext}}$, but converted to kpc based on distance to dwarf pair. }\\
\multicolumn{8}{l}{SFRs are derived from H$\alpha$ fluxes: a: \citet{whitney08}, b: \citet{wilke04}, c: \citet{gil03}  }\\
\multicolumn{8}{l}{d: \citet{koop04}, e: \citet{clemens02}, f: \citet{epinat08}, g: \citet{lanz13}, h: \citet{cox01}}\\
\multicolumn{8}{l}{i: \citet{kara04}}\\
\multicolumn{8}{l}{j: Defined as the diameter at which N(HI) = 1.2 $\times 10^{20}$ atoms cm$^{-2}$.  }
\end{tabular}
\end{table*}

%
%%-----------------------------------Host properties-----------------------------------

\begin{table*}
\caption{Properties of the host galaxies}
\label{table:hosts}
\begin{tabular}{cccccccc}
\hline
&Dwarf Pair &Host name& Proj. dist. host$^a$ & Host stellar mass& Host dark halo mass$^b$& $\Theta$$^c$ & Vel. sep., $\Delta$v$^d$\\
 &  & & [kpc] & $\times 10^{10}$[M$_{\odot}$]& $\times 10^{11}$[M$_{\odot}$] & & [km s$^{-1}$]\\
\hline
\bf{High tidal } &LMC, SMC &MW & 55   & 6.4 				&34	& 3.7&278 \\[2pt]
\bf{index }&IC 2058, PGC 75125 &NGC 1553 & 91   & 11.2 	&107	& 3.2 &298\\[2pt]
&NGC 4532, DDO 137 &NGC 4570 & 242   & 4.8 			&25	& 1.5 &224\\[2pt]

\\ \bf{Intermediate } &NGC 4618/25 &Messier 94& 242  & 3.4 	&20	& 1.4 &236\\[2pt]
\bf{tidal index }&NGC 4490/85  &NGC 4369 & 316  & 2.6	&	6.9& 0.91 &479\\[2pt]
&ESO435-IG16/IG20 &NGC 2997 &864 & 8.6 			&	60& 0.1 & 118 \\[2pt]

\\\bf{Isolated } &NGC 3448, UGC6016 & NGC 3310 & 1056 &2.6&	6.9 & -0.65 &355\\[2pt]
&UGC 9562/60 &none & $>$ 1500\\[2pt]
&NGC 672, IC 1727 & none & $>$ 1500 \\[2pt]
&NGC 4449, DDO 125 &none & $>$ 1500 \\[2pt]
\hline
\multicolumn{8}{l}{$^a$ The projected distances to the hosts are calculated based on the angular separation between the primary dwarf and the host galaxy at the distance of the  }\\
\multicolumn{8}{l}{primary (see Table \ref{tab:local}), after applying the velocity cuts (v$_{\text sep} < 500$ km s$^{-1}$ and D$_{\text{project}} < 1.5$ Mpc (except for the LMC/SMC where we used the average}\\
\multicolumn{8}{l}{  distance to the MW from the two galaxies).}\\
\multicolumn{8}{l}{$^b$ The dark matter halo masses are estimated using the \citet{moster13} abundance matching Eq. 2.}\\
\multicolumn{8}{l}{$^c$ $\Theta$ is the tidal index as defined in Equation \ref{index}. The pairs are sorted in order of decreasing $\Theta$.  }\\
\multicolumn{8}{l}{$^d$ Velocity separation of primary and host: $\Delta$v =  c $\times | z1 - z2 | /  ( 1 +  (z1 + z2)/2). $}\\

\end{tabular}
\end{table*}

%
%%-----------------------------------HI properties-----------------------------------

\begin{table*}
\caption{HI properties of dwarf pairs in our sample }
\label{table:HI}
\begin{tabular}{@{\:}l@{\:}c@{\:}c@{\:}c@{\:}c@{\:}c@{\:}c@{\:}c@{\:}c@{\:}}
\hline
Dwarf Pair &Beam size &Beam size & HI inner & HI total&HI total uni$^i$& NHI outer & Telescope &Bridge present\\ 
 &[arcsec]&[kpc]  & $\times 10^9$[M$_{\odot}$] & $\times 10^9$[M$_{\odot}$]&$\times 10^9$[M$_{\odot}$]& [10$^{19}$ atom/cm$^{2}$] & \\
 & &$\times (D/D_{\text{Table1}})$   & $\times (D/D_{\text{Table1}})^2$ & $\times (D/D_{\text{Table1}})^2$& $\times (D/D_{\text{Table1}})^2$ &  & \\
\hline
LMC/SMC& 930 &0.25& 				0.27/0.26&   	0.90	&	0.57	& 	0.2  & HIPASS$^b$ & yes  \\[2pt]
IC 2058/PGC 75125 & 60 &5.3&	 	 0.39/0.037&   	0.69	&	0.64	 & 4.5 &ATCA$^c$& yes \\[2pt]
NGC 4532/DDO 137 & 200& 13.4&	 	 0.28/0.068 &  	3.5 	&	1.3	& 0.1 & ARECIBO$^a$& yes \\[2pt]
NGC4618/25 & 60 &2.3&  			0.47/0.079 &  	1.4 	&	1.4	& 7.0 &WSRT$^e$& not clear \\[2pt]
NGC 4490/85  &  30& 1.0& 			2.4/0.23 & 	 3.7 	&	3.5	& 1.0 & VLA$^d$& yes \\[2pt]
ESO435-IG16/IG20 &130.4 &7.3&  		 0.14/0.055 & 	 0.8	&	0.69	 & 1.0 &ATCA$^e$& no  \\[2pt]
NGC 3448/UGC6016 & 60& 7.2& 		2.6/0.63 & 	 7.5 	&	7.5	& 7.0 &WSRT$^f$& yes \\[2pt]
UGC 9562/60 & 52.4 x 48.8 &6.5 x 6.0& 	0.2/0.07 & 	 2.0 	&	2.0	&7.0&VLA$^g$& yes \\[2pt]
NGC 672/IC 1727 & 60 &2.3&			1.1/0.42 & 	3.4 	&	3.4	& 7.0 & WSRT$^e$& yes \\[2pt]
NGC 4449/DDO 125 & 62 x 54 &1.1 x 1.0&  0.39/0.11& 	 1.1 	&	0.91	& 0.2 &VLA$^h$& no \\[2pt]
\hline
\multicolumn{9}{l}{Data from: a: \citealt{koop08}, b: \citealt{putman03}, c:  ATCA archives, d: \citealt{clemens98}, e: \citealt{hulst01}}\\
\multicolumn{9}{l}{f: \citealt{kim15}, g:  \citealt{cox01}, h: \citealt{hunter98}.}\\
\multicolumn{9}{l}{i: The total HI mass in the dwarf system after applying the uniform cut of N(HI) = $7 \times 10^{19}$ atoms cm$^{-2}$.}\\
\end{tabular}
\end{table*}

%%-----------------------------------------------------Bound vs unbound ----------------------------------------------------

\begin{table*}
\caption{Escape velocities as the edges of the HI profiles }
\label{tab:bound}
\begin{tabular}{cccccc}
\hline
Primary dwarf& Systemic velocity&HI extent distance & Dark mass of primary&Gas velocity at extent$^a$ &Escape velocity \\ 
name& [km s$^{-1}$] &[kpc]$\times (D/D_{\text{Table1}})$ &[$\times 10^{11}$M$_{\odot}$]& [km s$^{-1}$] & [km s$^{-1}$]\\ 
\hline
LMC 	& 	84 	& 	150	&	1.5 	&	326	&	123\\[2pt]
IC 2058 	&	1369 & 	9	&  	1.6	&	131	&	243\\[2pt]
NGC 4532&	2012 &  	150 	&	2.4 	&	242	&	155\\[2pt]
NGC 4618 &	533	&	14	&	2.0	&	56	&	249\\[2pt]
ESO435-IG16	&990	&	11	& 	1.5	&	20	&	230  \\[2pt]
NGC 4490 &	575	&	49	&	2.6	&	123	&	219\\[2pt]
NGC 3448 &	1350	&	30	&	1.8	&	150	&	210\\[2pt]
UGC 9562&	1292	&	20	&	1.4	&	58	&	204\\[2pt]
NGC 672&	429	&	20	&	2.1	&	121	&	241\\[2pt]
NGC 4449&	207	&	30	 &	1.8	&	93	&	212 \\[2pt]
\hline
\multicolumn{6}{l}{a: After subtracting systemic velocity of galaxy}\\
\end{tabular}
\end{table*}

%
%%-----------------------------------------------------Gunn & Gutt-----------------------------------------------------
\begin{table*}
\caption{Gunn \& Gott calculations}
\label{tab:cgm}
\begin{tabular}{cccccccc}
\hline
Dwarf galaxy& $R_\text{trunc}$$^a$&v$_\text{rot}$ & v$_\text{sep}$, $\Delta$v$^i$&$v_\text{esc}$&$\rho$ (v$_\text{sep}$)& $\rho$ (v$_\text{esc}$)& $r_t$$^j$ \\ 
name& [kpc] &[km s$^{-1}$] &[km s$^{-1}$] & [km s$^{-1}$]& $\times$10$^{-5}$ [cm$^{-3}$]& $\times$10$^{-5}$[cm$^{-3}$] & [kpc] \\
 & $\times (D/D_{\text{Table1}})$   &  & & & &  &$\times (D/D_{\text{Table1}})$ \\
\hline
LMC 	& 	5.2 & 	92$^b$&321 &	606&5.1&		1.4&		31\\[2pt]
SMC & 		4.2& 	60$^c$&217 &		587 &5.9&		0.81&		26\\[2pt]
IC 2058 	&	9.0 & 110$^d$&  298 &		857&4.2&		0.53&	50\\[2pt]
NGC 4532&	22.0 &  110$^e$ &224 &	382&7.8&		2.7&		132\\[2pt]
NGC 4618 &14&73$^f$&236&			347&2.2&		1.0&\\[2pt]
NGC 4490 &12&144$^g$&479&			194&29&		180&\\[2pt]
ESO435-IG16	&11& 40$^h$& 118&		383 &37&		3.5 &  \\[2pt]
\hline
\multicolumn{8}{l}{a: Defined as the radius at which the surface density profile in the direction towards the massive host deviates from the surface}\\
\multicolumn{8}{l}{ density in other directions. If no deviation is present, $R_{\text{trunc}}$ is defined as the extent of the data.}\\
\multicolumn{8}{l}{b: \citet{marel14}, c: \citealt{stani04}, d: our data, e: \citet{rubin99}, f: \citet{bush04}}\\
\multicolumn{8}{l}{g: \citet{huch80}, h: \citet{kim15}}\\
\multicolumn{8}{l}{i: From NED redshift of host and primary: $\Delta$v =  c $\times | z1 - z2 | /  ( 1 +  (z1 + z2)/2) $. For the LMC and SMC we use the 3D velocities}\\
\multicolumn{8}{l}{ (\citealt{kalli13}).}\\
\multicolumn{8}{l}{j: Tidal radius of high tidal index dwarfs defined in Equation \ref{rt}.}\\
\end{tabular}
\end{table*}

\nocite{*}
\bibliographystyle{mnras}
\bibliography{ms}

% Don't change these lines
\bsp	% typesetting comment
\label{lastpage}
\end{document}